\documentclass[usenatbib]{mn2e}
\usepackage{epsf,graphics,graphicx}
\usepackage{amsmath}
\usepackage{amssymb,latexsym,mathrsfs, bm}
\usepackage{color}
\usepackage{float}
\usepackage{natbib}
\usepackage{times}
\usepackage{caption}
\usepackage{subcaption}

\title[Neutrino mass and extra relativistic degrees of freedom in dark energy models]{Constraining neutrino mass and extra relativistic degrees of freedom in dynamical dark energy models using \emph{Planck} 2015 data in combination with low-redshift cosmological probes: basic extensions to \boldmath{$\Lambda$}CDM cosmology}
\author[Zhao, Li, Zhang \& Zhang]{Ming-Ming Zhao$^{1}$, Yun-He Li$^{1}$, Jing-Fei Zhang$^{1}$,
 Xin Zhang$^{1,2}$\thanks{Electronic address:
zhangxin@mail.neu.edu.cn}\\
$^{1}$ Department of Physics, College of Sciences, Northeastern University, Shenyang 110004,
China\\
$^{2}$ Center for High Energy Physics, Peking University, Beijing 100080, China}

\begin{document}
\date{ }
\maketitle

\begin{abstract}
We investigate how the properties of dark energy affect the cosmological measurements of neutrino mass and extra relativistic degrees of freedom. We limit ourselves to the most basic extensions of $\Lambda$ cold dark matter (CDM) model, i.e. the $w$CDM model with one additional parameter $w$, and the $w_{0}w_{a}$CDM model with two additional parameters, $w_{0}$ and $w_{a}$. In the cosmological fits, we employ the 2015 cosmic microwave background temperature and polarization data from the \emph{Planck} mission, in combination with low-redshift measurements such as the baryon acoustic oscillations, Type Ia supernovae and the Hubble constant ($H_{0}$). Given effects of massive neutrinos on large-scale structure, we further include weak lensing, redshift space distortion, Sunyaev--Zeldovich cluster counts and \emph{Planck} lensing data. We show that, though the cosmological constant $\Lambda$ is still consistent with the current data, a phantom dark energy ($w<-1$) or an early phantom dark energy (i.e. quintom evolving from $w<-1$ to $w>-1$) is slightly more favoured by current observations, which leads to the fact that in both $w$CDM and $w_0w_a$CDM models we obtain a larger upper limit of $\sum m_\nu$. We also show that in the three dark energy models, the constraints on $N_{\rm eff}$ are in good accordance with each other, all in favour of the standard value 3.046, which indicates that the dark energy parameters almost have no impact on constraining $N_{\rm eff}$. Therefore, we conclude that the dark energy parameters can exert a significant influence on the cosmological weighing of neutrinos, but almost cannot affect the constraint on dark radiation.

\end{abstract}



\section{Introduction}\label{intro}

Since the phenomenon of neutrino oscillation was revealed by the solar and atmospheric neutrino experiments, the facts that neutrinos have masses and there is a significant mixing between different neutrino species have been convincingly confirmed. However, it is a great challenge for particle physics experiments to directly measure the absolute neutrino mass scale. In fact, the neutrino oscillation experiments are only sensitive to the squared mass differences between the neutrino mass eigenstates. The current data from the solar and atmospheric neutrino experiments give $\Delta m_{21}^2\simeq 7.6\times 10^{-5}$ eV$^2$ and $|\Delta m_{32}^2|\simeq 2.4\times 10^{-3}$ eV$^2$ \citep{Olive2014}, respectively. These measurements give rise to two possible mass orders, i.e. the normal hierarchy with $m_1<m_2\ll m_3$ and the inverted hierarchy with $m_3\ll m_1<m_2$.

To work out the absolute masses of neutrinos, one needs at least an additional relationship between the three neutrino mass eigenstates. The neutrino oscillation measurements can only provide a lower limit for the sum of the neutrino masses, $\sum m_\nu\gtrsim 0.06$ eV. Actually, particle physics experiments can also measure the total mass of neutrinos, but these experiments are fairly difficult. For example, the tritium beta-decay experiments, i.e. Troitsk and Mainz, gave an upper bound, $m_\beta<2.3$ eV (95 per cent confidence level), where $m_\beta$ is a mass to which the beta-decay experiments are sensitive \citep{Kraus2005,Otten2008}. The KATRIN (KArlsruhe TRItium Neutrino) experiment aims to measure $m_\beta$ with a sensitivity of $\sim0.2$ eV, which would give an upper bound for the total mass, $\sum m_\nu<0.6$ eV \citep{KATRIN Collaboration2001,Wolf2010}. In addition, the neutrinoless double beta decay ($0\nu\beta\beta$) experiments would also measure the effective mass of Majorana neutrinos at the level of ${\cal O}$(0.1--1) eV depending on the mixing matrix \citep{Klapdor-Kleingrothaus2001,Klapdor-Kleingrothaus2004}. However, compared to the particle physics experiments, it has been found that the cosmological observations are actually more prone to be capable of measuring the absolute neutrino mass \citep{legourgues2006,Valle2006,Hannestad2011,legourgues2012}. Massive neutrinos could leave distinct signatures on the cosmic microwave background (CMB) and large-scale structure (LSS) at different epochs of the Universe's evolution \citep{Abazajian2014}. To a large extent, these signatures could be extracted from the available cosmological observations, from which the total neutrino mass could be constrained. Currently, the CMB power spectrum, combined with LSS and cosmic distance measurements, can provide tight limits on the total mass of neutrinos \citep{planck2013,planck2015a}.

The CMB temperature and polarization power spectra from \emph{Planck} 2015 in combination with the baryon acoustic oscillations (BAO) data give a 95 per cent limit of $\sum m_{\nu} < 0.17$ eV based on the $\Lambda$ cold dark matter (CDM) model \citep{planck2015a}. This constraint depends much on the effect of massive neutrinos on the CMB power spectrum and BAO measurements at low redshfts. At low redshifts, neutrinos are non-relativistic, and they contribute to the expansion rate through matter density, and thus change the angular diameter distance $D_{A}$. Further, the acoustic peak scale of CMB power spectrum and distance $D_{V}$ of hybrid quantity $r_{s}(z_{\rm drag})/D_{V}$ measured by BAO are altered by the changed $D_{A}$. Besides, massive neutrinos can also affect the CMB power spectrum through the integrated Sachs--Wolfe (ISW) effect \citep{Hall2012,legourgues2012,Hou2014}. Increasing neutrino mass leads to the decay of gravitational potential inside the Hubble radius. As photons go through this decaying potential on their way towards observer, new anisotropies are generated by the late ISW effect. Moreover, at the period when neutrinos transform from relativistic to non-relativistic regime, they also lead the gravitational potential to decay and the new anisotropies are generated by the early ISW effect  \citep{Kaplinghat2003,legourgues2006}.

Massive neutrinos can also leave key signatures in the spectrum of matter fluctuations by the absence of neutrinos perturbations in matter power spectrum, and hence in some large-scale observations \citep{Bond1997, legourgues2006}.
For example, the weak gravitational lensing provides a potentially powerful measurement of the amplitude of matter spectrum at low redshifts with cosmic shear.
The matter fluctuation spectrum is also related to the growth factor $D(z)$. The redshift space distortion (RSD) can offer a direct measurement for growth rate, $f(z)$, at several low redshifts, where $f(z)= d\ln D/{d}\ln a$.
Recently, the cluster abundance extracted from the \emph{Planck} Sunyaev--Zeldovich (SZ) catalogue is considered, which depends on measurements for the amplitude of the density perturbations today,
characterized by the equivalent linear theory extrapolation, the root-mean-square mass fluctuation, $\sigma_{8}$. Since massive neutrinos suppress the lensing power, the CMB lensing is also helpful for constraining the neutrino mass.

On the other hand, the cosmological measurements also allow for constraining the extra relativistic degrees of freedom, parametrized via $N_{\rm eff}$, usually called dark radiation. In the standard model, we have $N_{\rm eff}=3.046$ \citep{Mangano2005}. A variation in $N_{\rm eff}$ can also affect the CMB power spectrum through a few ways, for example, changing the redshift of the matter-radiation equality, impacting on the amplitude of the peaks at high multipoles, and the early ISW effect. Therefore, $N_{\rm eff}$ can be constrained by the CMB power spectrum \citep{Bashinsky2004,Smith2011,Archidiacono2013}. In the last few years, there have been some mild preference for a non-standard value of the extra relativistic degrees of freedom from the CMB anisotropy measurements \citep{Dunkley2011,Keisler2011,Komatsu2011,Hinshaw2013,Hou2014}.
However, the recent high-precision CMB temperature spectrum from \emph{Planck} leads to evidence for a standard value of $N_{\rm eff}$ \citep{planck2015a}. The effective number of relativistic species in the Universe is not clear yet, which also needs the inclusion of other astronomical direct measurements to have a cosmological probe for it. In cosmology, the total relativistic energy density in neutrinos and any other dark radiation is given in terms of the photon density $\rho_{\gamma}$ by $\rho=N_{\rm eff}(7/8)(4/11)^{4/3}\rho_{\gamma}$.

Usually, the cosmological constraints on $\sum m_{\nu}$ and $N_{\rm eff}$ are derived, based on the standard $\Lambda$CDM cosmology. In much more complicated models, using the CMB power spectrum is possible to accommodate different neutrino mass or dark radiation \citep{Hannestad2005,zhang2015b,zhang2016}. In this paper, we will consider the basic extensions of the $\Lambda$CDM model, i.e. the $w$CDM and $w_{0}w_{a}$CDM models \citep{Chevallier2001,Linder2006,Linder2008}, in which we wish to provide the simplest examples that how the dark energy property affects the cosmological weighing of neutrinos. Exploration of the effects of the dark energy property on the neutrino mass bound is based on the fact that dark energy can have effects on the CMB power spectrum through changing the acoustic peak scale, the late ISW effect and so on \citep{planck2015gravi}, while the effects can also be caused by massive neutrinos. It is known that the constraints on the neutrino mass will be sensitive to the dark energy property when the CMB power spectrum are utilized to have a cosmological measurement for them. To see clearly how the cosmological measurements of neutrino mass is affected by the dark energy property, we will focus on the constraints on the neutrino mass in the $w$CDM and $w_{0}w_{a}$CDM models from the CMB power spectrum. For our previous studies on this aspect, see e.g. \cite{zhang2016} and \cite{Wang2016}. But the cases of dark radiation in the dynamical dark energy models are not addressed in these previous works. As a supplement, we will also concentrate upon the constraints on dark radiation in these scenarios.

In the global fitting, the addition of the dynamical dark energy will increase the degeneracies in the cosmological parameters, and thus using the CMB power spectrum alone is not enough. We need to combine some geometric observations,
for example, the BAO data, the Type Ia supernova (SN) data, and the independent measurement of Hubble constant ($H_{0}$). Here, BAO, SN and $H_{0}$ can break the degeneracies at the low redshifts, and they can provide strong exploration to the equation of state (EoS) of dark energy at $z\lesssim1$. Here, to constrain the neutrino mass well, we will also use the LSS observations, including the WL, RSD, SZ and CMB lensing data.

In fact, there has been a large number of work on the issue of investigating neutrino mass and dark radiation using cosmological observations in the literature. For example, \cite{Hou2013} presented the effects of $N_{\rm eff}$ on the CMB peaks; \cite{Serra2007} discussed forecasted constraints on massive neutrinos and dark energy; \cite{Santos2013} presented constraints on massive neutrinos and dark energy with reference to clustering (but also contained physical descriptions); \cite{Giusarma2013} presented constraints on massive neutrinos and dark energy after the first release of \emph{Planck} mission. In addition, \cite{MacCrann2015} discussed a variety of combinations to address the possible discordance between the \emph{Planck} constraints and low$-$redshift probes. \cite{Di Valentino2015} considered a 12-parameter extended cosmological model that allows for dark energy, massive neutrinos and dark radiation, simultaneously; see also \cite{Di Valentino2016}. In particular, \cite{zhang2016} and \cite{Wang2016} recently considered constraints on the neutrino mass in the $w$CDM model and the holographic dark energy model (without and with the consideration of mass hierarchies, respectively).

Therefore, under such circumstances, one might be concerned with the primary aim of this paper. Here, we briefly discuss the basic motivations of this work. (i) We wish to use the latest cosmological observations to constrain the neutrino mass $\sum m_\nu$ and the dark radiation parameter $N_{\rm eff}$ in the basic extensions to $\Lambda$CDM cosmology, i.e. the $w$CDM model and the $w_0w_a$CDM model. We will obtain the new constraint results of $\sum m_\nu$ and $N_{\rm eff}$ as well as $w$ and $(w_0, w_a)$ and other parameters using different combinations of current observational data sets, which is a useful reference to other relevant studies. (ii) We wish to make a uniform comparison for the results in the $\Lambda$CDM, $w$CDM and $w_0w_a$CDM models. Here, we use totally the same data sets to do the comparison analysis. From the uniform comparison, we will see how the constraint results change when varying cosmological models and data combinations. (iii) We wish to investigate how the dark energy parameters affect the cosmological constraints on the neutrino mass and dark radiation in the basic extensions of $\Lambda$CDM. Through our analysis in depth and in detail, we will see that the dark energy parameters can exert a significant influence on the constraints of $\sum m_\nu$, but almost cannot affect the constraints on $N_{\rm eff}$.

The paper is organized as follows. In Section~\ref{sec:method}, we describe the observations we use in this paper. In Section~\ref{sec:results}, we present the constraint results of the neutrinos mass in the $\Lambda$CDM, $w$CDM and $w_{0}w_{a}$CDM models. In Section~\ref{radiations}, we also present the constraints on dark radiation in the models mentioned above. Finally, we give conclusions in Section~\ref{sec:conclusion}.

\section{methodology and data }\label{sec:method}
In our analysis, we allow for the inclusion of $\sum m_{\nu}$ or $N_{\rm eff}$ in the $\Lambda$CDM, $w$CDM and $w_{0}w_{a}$CDM models. The cosmological parameters in the base $\Lambda$CDM model are
\begin{flushleft}
\begin{equation}\label{2}
   \{\Omega_{\rm b}h^{2}, \Omega_{\rm c}h^{2}, 100\theta_{\rm MC}, \tau, n_{\rm s}, \log[10^{10}A_{\rm s}]\},
\end{equation}
\end{flushleft}
where $\Omega_{\rm b}h^{2}$ is baryon energy density, $\Omega_{\rm c}h^{2}$ is the CDM energy density, $100\theta_{\rm MC}$ is 100 times the ratio between the sound horizon and the angular diameter distance at the decoupling, $\tau$ is the reionization optical depth and $n_{s}$ and $A_{s}$ are the primordial spectral index and the amplitude of the primordial spectrum, respectively.
There are extra parameters in the cosmological global fittings when considering massive neutrinos or dark radiation in the $w$CDM and $w_{0}w_{a}$CDM models.
The extra cosmological parameters include $\sum m_\nu$, $N_{\rm eff}$, $w$, $w_0$ and $w_a$, in our analysis. We constrain the above cosmological parameters with three data combinations.

First, our baseline combination is comprised of CMB measurements and BAO data. The CMB measurements include the full \emph{Planck} 2015 release of TT temperature spectrum and TE and EE polarization spectra at whole multipoles ($2<\ell<2900$) \citep{Aghanim2015}, and we refer to this combination as `Planck' in our work. We use the BAOs data in good agreement with the \emph{Planck} data, including the measurements from the 6dFGS ($z_{\rm eff} = 0.1$) \citep{Beutler2011}, SDSS-MGS ($z_{\rm eff} = 0.15$) \citep{Ross2014}, LOWZ ($z_{\rm eff} = 0.32$) and CMASS ($z_{\rm eff} = 0.57$) samples of BOSS  \citep{Anderson2014}. This combination is usually denoted as `Planck+BAO'.

Secondly, to further constrain the properties of dark energy, we consider geometric measurements at low redshifts, including the Type Ia SN observation and the direct measurement of the Hubble constant. For the Type Ia SN observation, we employ the `Joint Light-curve Analysis' (JLA) sample \citep{Betoule2014}, compiled from the SNLS, SDSS and the samples of several low-redshift SNe. For the local measurement of the Hubble constant, we employ the result of \citet{Efstathiou2014}, derived from a re-analysis of the Cepheid data of \citet{Riess2011}, with the measurement value $H_{0}=70.6\pm3.3\rm \,km\, s^{-1} \rm Mpc^{-1}$. We denote the above data as SN and $H_{0}$ in the data combinations.

Thirdly, we also consider measurements from the growth of structure to constrain the neutrino mass, including weak lensing, RSDs, SZ cluster counts and CMB lensing. For the weak lensing data, we use the cosmic shear data provided by the CFHTLenS survey \citep{Heymans2012,Erben2013}, which perform tomographic analysis with cosmological cuts, specifically removing the angular scales $\theta < 3$ arcmin for two lowest bin combination, angular scales $\theta < 30$ arcmin for $\xi^{-}$ for four lowest bins, and $\theta<16$ arcmin for two highest bins for $\xi^{+}$. We denote this measurement as WL. The RSD data provide powerful constraints on the growth rate of structure by measuring the parameter combination $f\sigma_{8}(z)$. Here, we follow the results of \citet{Samushia2014}, employing the covariance matrix for the three parameters, $D_{\rm v}/r_{\rm drag}$, $F_{\rm AP}$ and $f\sigma_{8}$. It should be noticed that one data point is repeatedly used for BAO at $z=0.57$. Therefore, we exclude the BOSS CMASS result from BAO when using two measurements simultaneously. Then, for the SZ cluster counts, we use the full mission data from \emph{Planck} with a larger catalogue of SZ clusters \citep{planck2015SZ}, which still keeps the overall mass bias characterized by $1-b$ parameter, varied in a prior with [0.1,~1.3] range. We describe this measurement as SZ. Finally, for CMB lensing, we use the \emph{Planck} lensing measurement \citep{planck2015lensing}.

Our constraints are based on the latest version of the Monte Carlo Markov Chain package {\small COSMOMC} \citep{Lewis2002} and we perform the method of $\chi^{2}$ statistic in the calculations.

\section{Constraints on neutrino mass}\label{sec:results}
In this section, we investigate the constraints on the total neutrino mass $\sum m_\nu$ in dynamical dark energy models. As mentioned above, for dynamical dark energy models, we only consider the basic extensions to $\Lambda$CDM, i.e. the $w$CDM model and the $w_0w_a$CDM model.

We use three data combinations to do the analysis, that are Planck+BAO, Planck+BSH and Planck+BSH+LSS. Here, for convenience, we use `BSH' to denote the joint BAO+SN+$H_0$ data and use `LSS' to denote the joint WL+RSD+SZ+lensing data.

It has been shown by \cite{planck2015a} that the \emph{Planck} data are consistent with the BAO, SN (JLA compilation) and $H_0$ (the result of \citealt{Efstathiou2014}) data, and thus the data consistency for the combinations of Planck+BAO and Planck+BSH can be ensured. But, it is also known that the LSS data prefer a low value of $\sigma_8$ compared to the \emph{Planck} fitting result based on $\Lambda$CDM. Here, we note that (i) we have carefully, conservatively use the WL and RSD data totally according to the prescription of the Planck Collaboration \citep{planck2015a}, and (ii) once considering the massive neutrinos and dynamical dark energy in the cosmological model, the tension between Planck and LSS can be greatly relieved \citep{limiao2013,Battye2014,Wyman2014,zhangjingfei2014,zhangjingfei2015}. Thus, it is also reasonable to use the combination of Planck+BSH+LSS in this work.

In the following, we will first present the effects of massive neutrinos and dark energy on the observations, in particular the CMB observation, and then use the actual observations to constrain these parameters.

\subsection{Effects of massive neutrinos and dynamical dark energy on CMB temperature spectrum}\label{caused}

The CMB observation could provide an accurate measurement of the angular diameter distance $D_A$ to last-scattering surface with the redshift $z_{*}\simeq 1100$, which is rather important for constraining cosmological parameters because a precise high-redshift measurement could play a significant role in determining the whole expansion history.
The angular diameter distance $D_A$ is linked to the expansion history of the Universe through the relation
\begin{equation}\label{distance}
  D_{A}(z)=\frac{1}{1+z}\int^{z}_{0}\frac{{d}z'}{H(z')}.
\end{equation}

We first discuss the effects of dynamical dark energy on the CMB observation. For simplicity, we consider $w$CDM as an example, i.e. we assume $w$ is a constant. Increasing $w$ at fixed matter density increases $H(z)$ at $z\lesssim1$ and reduces the angular diameter distance $D_A(z)$ for $0<z\leq z_{*}$ \citep{Howlett2012}. The observable, $\theta_{*}=r_{s}/D_{A}$, determines the acoustic peak scale of CMB power spectrum, and as a consequence, the reduced $D_A(z)$ increases the peak of the CMB spectrum.

Another main effect of dark energy on CMB power spectrum is from the late ISW effect. Before the dark energy domination, the gravitational potential in the Universe keeps as a constant to the first order in linear perturbation theory. When dark energy starts to dominate the Universe's evolution, the gravitational potential is not a constant any more, due to the accelerated expansion of the Universe. Dark energy leads to the decay of gravitational potential on large scales, generating new anisotropies of the CMB photons. As the photons go through these decaying potentials on their way towards the observer, new anisotropies are generated by the ISW effect. Eventually, the small-scale CMB anisotropy spectrum is actually also altered.

\begin{figure*}
\includegraphics[scale=0.3, angle=0]{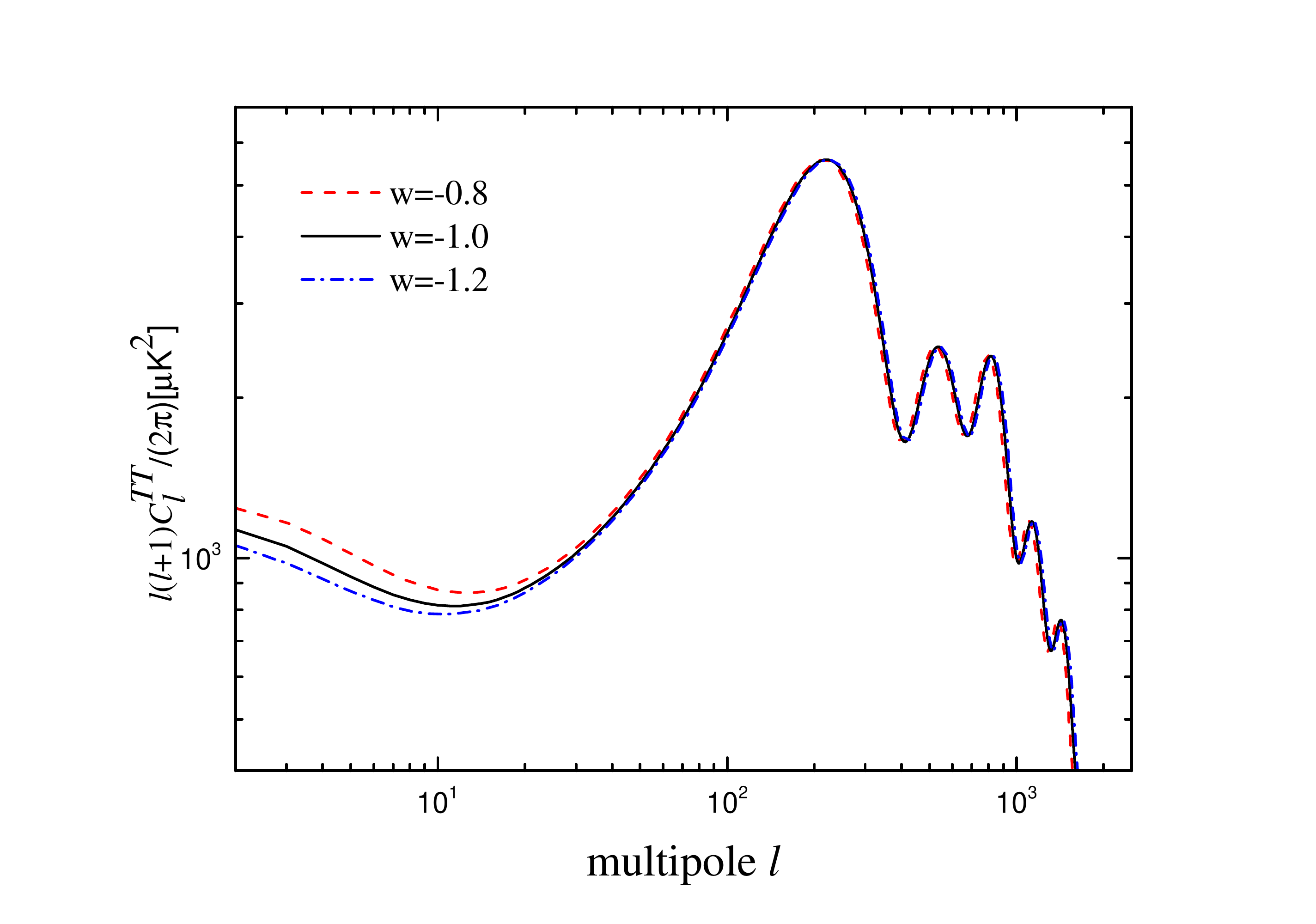}
\includegraphics[scale=0.3, angle=0]{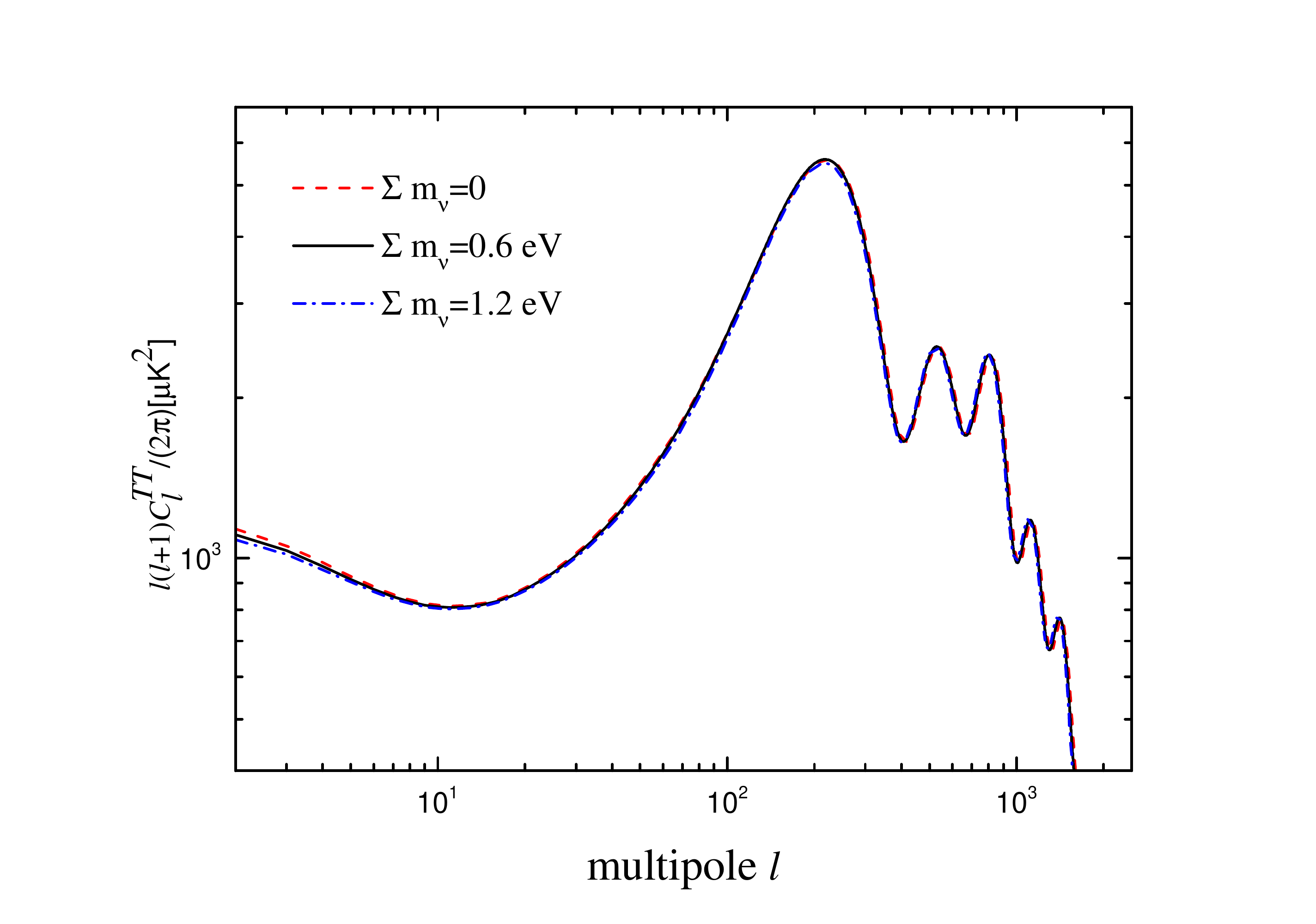}
\caption{Left-hand panel: The CMB temperature spectra $C_{\ell}^{TT}$ with different EoS of dark energy $w$. Here, we choose $w$ $=$ $-0.8$, $-1.0$ and $-1.2$, and fix $\sum m_{\nu}$ to be 0.06 eV. At $2$ $<$ $\ell$ $<$ $50$, it is found that a smaller $w$ leads to a suppression of CMB temperature power, namely, a smaller $C_{\ell}^{TT}$, due to the late ISW effect. Right-hand panel: the CMB temperature spectra with different total neutrino mass $\sum m_{\nu}$. Here, we choose $\sum m_{\nu}$ $=$ $0$, $0.6$ eV and $1.2$ eV, and fix $w$ $=$ $-1$. At $2$ $<$ $\ell$ $<$ $50$, a larger $\sum m_{\nu}$ leads to a smaller $C_{\ell}^{TT}$, due to the late ISW effect.}
\label{fig:CMBpoer}
\end{figure*}

The left-hand panel of Fig.~\ref{fig:CMBpoer} shows how $C_{\ell}^{TT}$ changes with different EoS of dark energy $w$. Here, as examples, we choose three values of $w$, namely, $w=-0.8$, $-1.0$ and $-1.2$, and fix $\sum m_{\nu}$ to be 0.06 eV and other parameters consistent with \emph{Planck} \citep{planck2015a}. The figure shows that, at the low multipoles ($2<\ell<50$), a smaller $w$ leads to a suppression of temperature spectrum, i.e. a smaller $C_{\ell}^{TT}$, due to the late ISW effect.

Massive neutrinos can also affect the CMB power spectrum, through altering expansion rate and gravitational potential \citep{Hu2002,Ichikawa2005}.
Initially, neutrinos are massless and behave as radiation. After recombination, massless neutrinos generally transform to massive neutrinos. In this period, neutrinos are non-relativistic, but they still contribute to energy density. However, this behaviour is neglected in the Poisson equation. As a consequence, the gravitational potential decays through the increased $H(z)$. As photons free stream immediately after decoupling, the anisotropies are created by the early ISW effect. Thus, the neutrino mass affects the CMB power spectrum.

When neutrinos are absolutely non-relativistic, they contribute to the expansion rate through the matter density. Increasing neutrino mass leads to an reduction in $H(z)$ at $z\lesssim1$ at fixed $\theta_{*}$. The decreased $H(z)$ results in a decay of gravitational potential at small scales, and thus contributes to a suppression of CMB power spectrum through the late ISW effect \citep{Hou2014}. In addition to that, the decreased $H(z)$ increases $D_{A}$. 

The right-hand panel of Fig.~\ref{fig:CMBpoer} shows how CMB temperature spectrum changes with different neutrino mass $\sum m_\nu$. Here, for example, we choose three values of $\sum m_{\nu}$, namely, $\sum m_\nu=0$, 0.6 eV and 1.2 eV, and we fix $w$ to be $-1$. We find that larger neutrino masses suppress CMB spectrum at low multipoles ($2<\ell<50$) due to the late ISW effect.

\subsection{Massive neutrinos versus dark energy}

We now focus on the constraints on the neutrino mass in the $w$CDM and $w_{0}w_{a}$CDM models from the above mentioned three data combinations. According to the constraint results, we investigate the correlation between neutrino mass and dark energy parameter, from which we can see how the property of dark energy impacts on the cosmological measurement of neutrino mass.

\begin{figure}

\begin{center}
\includegraphics[scale=0.9, angle=0]{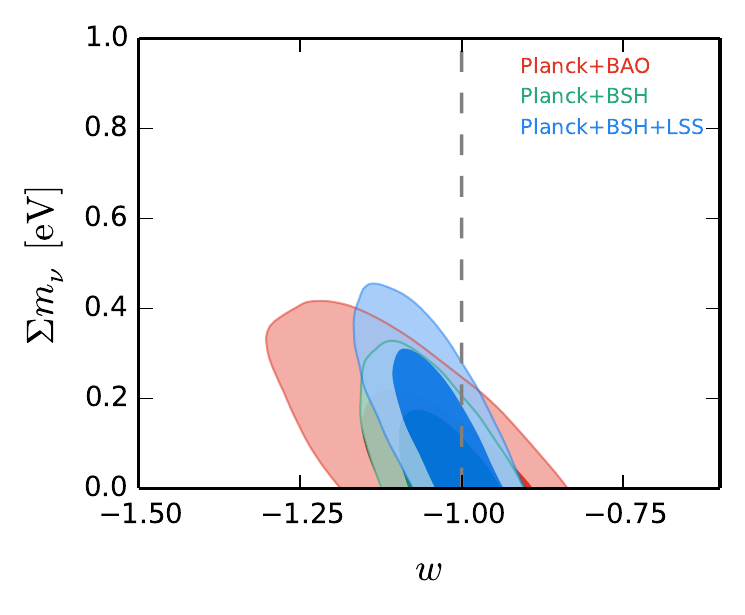}
\caption{68 per cent and 95 per cent CL contours in the $w$$-$$\sum m_{\nu}$ plane from the three data combinations of Planck+BAO, Planck+BSH and Planck+BSH+LSS, where `BSH' denotes the joint of BAO, SN and $H_{0}$ data, `LSS' denotes the combination of WL, RSD, SZ and CMB lensing data. Planck+BSH gives a tighter constraint on $\sum m_{\nu}$, but Planck+BSH+LSS allows a larger $\sum m_{\nu}$. The constraints on dark energy from the three data combinations are all compatible with $\Lambda$CDM.}
\label{fig:w-mnu}
\end{center}
\end{figure}

The marginalized posterior contours in the $\sum m_{\nu}$--$w$ plane for the $w$CDM model is shown in Fig.~\ref{fig:w-mnu}. In this figure, the three data combinations give consistent results, showing that $w$ is anti-correlated with $\sum m_{\nu}$. This correlation can be explained by the compensation to the effects on the acoustic peak scale $\theta_{*}$. Increasing $w$ leads $H(z)$ to increase. However, a reduction in $\sum m_{\nu}$ can compensate the changed $H(z)$, and there are the same $D_{A}$ according to equation (\ref{distance}) and the same $\theta_{*}$. The results show that a larger neutrino mass is allowed by a phantom dark energy in $w$CDM. The Planck+BSH combination provides the tightest constraint on $\sum m_\nu$. Once the LSS data are added, the constraint becomes looser. This is because the current LSS observations favour lower matter perturbations (demonstrated by a lower $\sigma_8$), which obviously tends to favour a larger neutrino mass due to the free-streaming effect of massive neutrinos.

\begin{figure}
\begin{center}
\includegraphics[scale=0.4, angle=0]{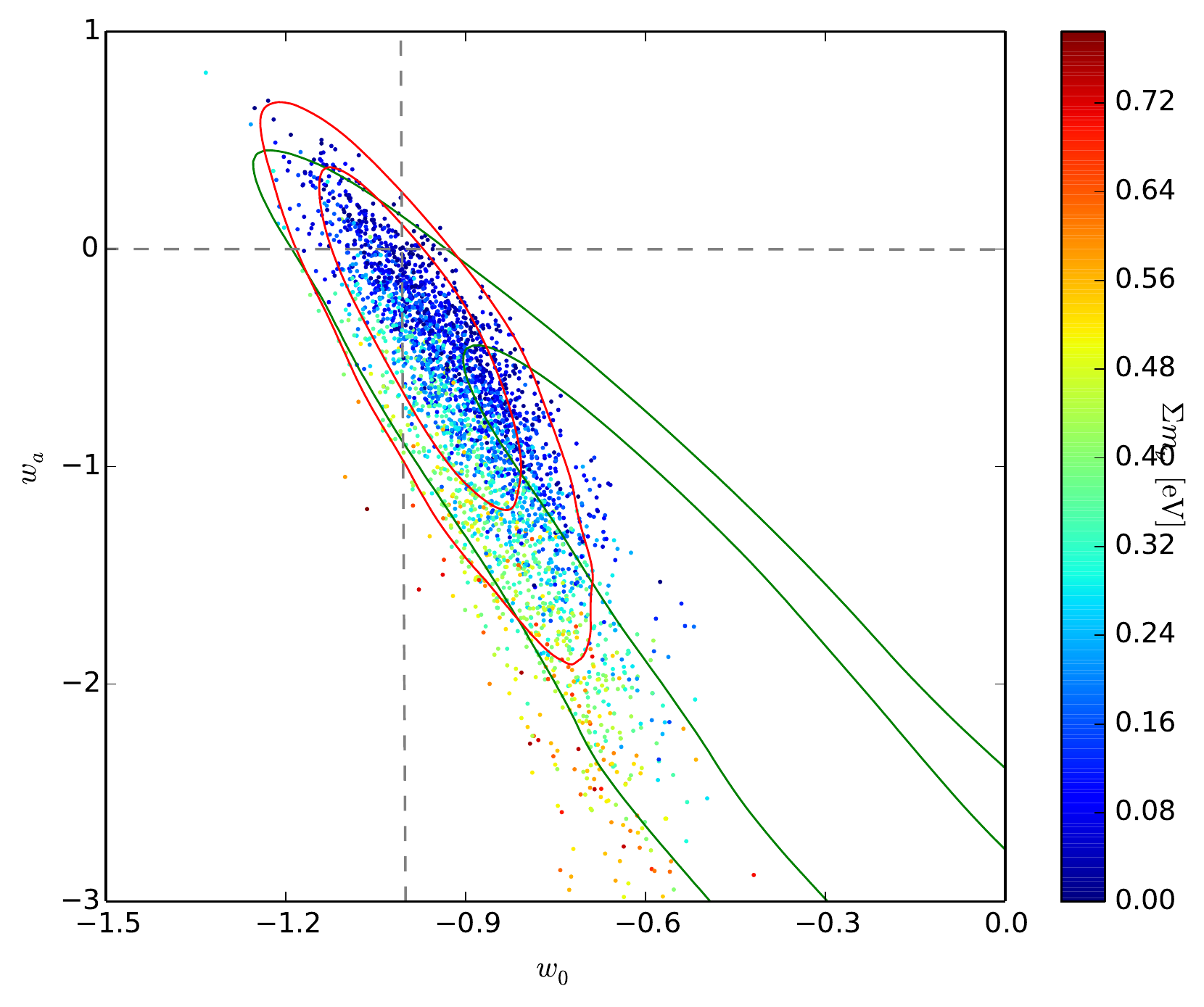}
\caption{Samples from the Planck+BSH chains in the $w_{0}$$-$$w_{a}$ plane, colour-coded by $\sum m_{\nu}$. The green contours show the constraints from the Planck+BAO data set, and the red contours show the constraints from Planck+BSH+LSS.
The $\Lambda$CDM case with $w_{0}$ $=$  $-1$ and $w_{a}$ $=$ $0$ is shown in the plane by the cross of horizontal and vertical dashed lines. The samples show the points corresponding to larger $\sum m_{\nu}$ distribute the regions of $w$ evolving from $w$ $<$ $-1$ to $w$ $>$ $-1$.}
\label{fig:w0-mnu}
\end{center}
\end{figure}

For the case of the $w_0w_a$CDM model, we plot in Fig.~\ref{fig:w0-mnu} the marginalized posterior contours in the $w_0$--$w_a$ plane from the Planck+BAO and Planck+BSH+LSS combinations, shown as the green and red contours, respectively. The constraints in the $w_0$--$w_a$ plane from the Planck+BSH data are also shown in Fig.~\ref{fig:w0-mnu} as samples, colour coded by the value of $\sum m_\nu$. From this figure, we find that a larger $\sum m_\nu$ is favoured by an early phantom dark energy, more precisely, a dynamical dark energy evolving from $w<-1$ to $w>-1$.

Next, we compare the constraint results of $\sum m_\nu$ for the $\Lambda$CDM model, the $w$CDM model and the $w_{0}w_{a}$CDM model. The detailed fitting results are given in Tables \ref{tab1} and \ref{tab2}.

The Planck+BAO data combination gives the limits: $\sum m_{\nu}<0.17$ eV (95 per cent CL) for $\Lambda$CDM (in exact agreement with the result derived by the Planck Collaboration), $\sum m_{\nu}<0.33$ eV (95 per cent CL) for $w$CDM and $\sum m_{\nu}<0.47$ eV (95 per cent CL) for $w_{0}w_{a}$CDM. For $w$CDM, we have $w=-1.068^{+0.103}_{-0.067}$, and for $w_{0}w_{a}$CDM, we have $w_{0}=-0.52^{+0.34}_{-0.22}$ and $w_a=-1.73^{+0.39}_{-1.25}$. We find that, compared to $\Lambda$CDM, the upper limits of neutrino mass become larger in the $w$CDM and $w_{0}w_{a}$CDM models. In the case of dynamical dark energy, we can clearly see that a phantom energy with $w<-1$ is more favoured and a quintom energy evolving from $w<-1$ to $w>-1$ is more favoured by the current observations, and thus a larger $\sum m_\nu$ is more favoured in the two dynamical dark energy models compared to $\Lambda$CDM.

In the joint fits to Planck+BSH, we obtain $\sum m_{\nu}<0.15$ eV for $\Lambda$CDM, $\sum m_{\nu}<0.25$ eV for $w$CDM and $\sum m_{\nu}<0.51$ eV for $w_{a}w_{0}$CDM. Correspondingly, we have $w=-1.042^{+0.052}_{-0.045}$ for $w$CDM and we have $w_{0}=-0.89^{+0.12}_{-0.14}$ and $w_{a}=-0.84^{+0.80}_{-0.49}$ for $w_{0}w_{a}$CDM. We find that, in this case, the EoS of dark energy can be constrained more tightly because of the addition of SN and $H_0$ data. For the $\Lambda$CDM and $w$CDM models, the Planck+BSH data give distinctly tighter constraints on $\sum m_\nu$ than the Planck+BAO data. For the $w_0w_a$CDM model, the constraint of $\sum m_\nu$ becomes a little bit looser for Planck+BSH than for Planck+BAO.

Under the Planck+BSH+LSS data, the total mass of neutrinos are constrained to $\sum m_{\nu}<0.22$ eV for $\Lambda$CDM, $\sum m_{\nu}<0.36$ eV for $w$CDM and $\sum m_{\nu}<0.52$ eV for $w_{0}w_{a}$CDM. Correspondingly, we have $w=-1.042^{+0.057}_{-0.047}$ for $w$CDM and we have $w_{0}=-0.96\pm0.11$ and $w_{a}=-0.47^{+0.59}_{-0.43}$ for $w_{0}w_{a}$CDM. We find that, compared to the case of Planck+BSH, the addition of the LSS data only makes little improvement to the constraints on dark energy. This is because the smooth dark energy affects the growth of structure only through the expansion history and thus the measurements of matter perturbations can only provide loose constraints on the property of dark energy, especially for the case that the current measurements of growth of structure are not accurate enough. But we find that, by adding the LSS data, the constraints on $\sum m_\nu$ become looser, compared to the Planck+BSH case. As mentioned above, the current LSS observations, such as WL, RSD and SZ, prefer a Universe with low matter perturbations, compared with the \emph{Planck} CMB data \citep{planck2015a}. The tension between \emph{Planck} and LSS can be greatly relieved by considering massive neutrinos in the cosmological model due to the free-streaming effect of massive neutrinos tending to suppress the matter perturbations \citep{Battye2014,Wyman2014,zhangjingfei2014,zhangjingfei2014a,zhangjingfei2015}. Therefore, a larger $\sum m_\nu$ is allowed when the LSS data preferring low matter perturbations are included.

\begin{figure}
\begin{center}
\includegraphics[scale=0.25, angle=0]{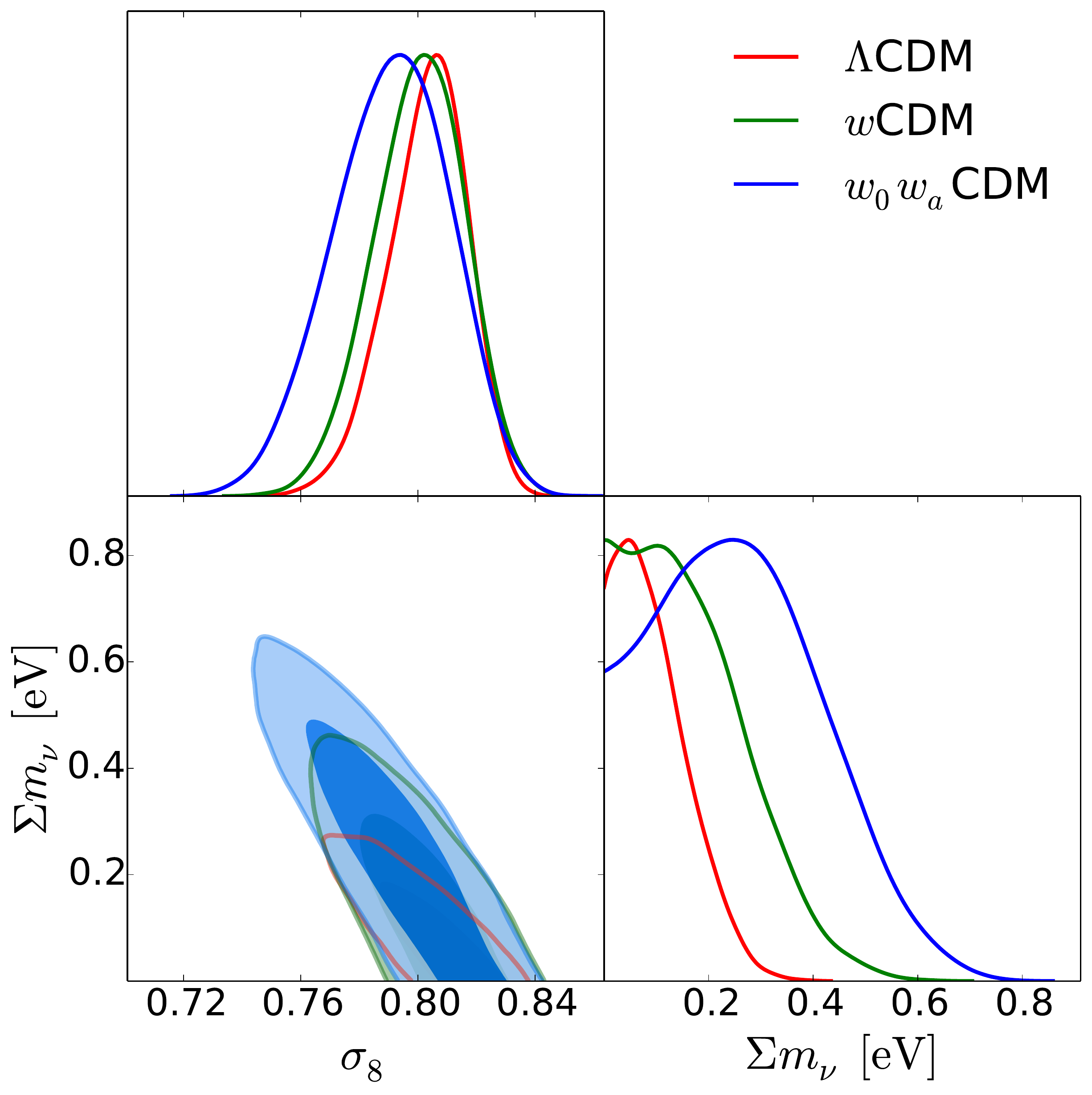}
\caption{Joint, marginalized constraints from Planck+BSH+LSS on the $\Lambda$CDM (red), $w$CDM (green) and $w_{0}w_{a}$CDM (blue) models. The 68 per cent and 95 per cent CL contours in the $\sigma_{8}$$-$$\sum m_{\nu}$ plane are shown. Note here that there is a peak in the posterior distribution of $\sum m_{\nu}$ for the $w_{0}w_{a}$CDM case around $\sum m_{\nu}$ $=$ $0.285$ eV, but the statistical significance is rather low.}
\label{fig:HWDE}
\end{center}
\end{figure}

Fig. \ref{fig:HWDE} shows the joint, marginalized constraints on $\sum m_\nu$ and $\sigma_{8}$ for the $\Lambda$CDM, $w$CDM and $w_0w_a$CDM models from the data combination of Planck+BSH+LSS. We find that, in all the three models, $\sigma_{8}$ is indeed anti-correlated with $\sum m_\nu$. Thus, in a cosmological model, considering massive neutrinos can lead to a low $\sigma_8$ Universe, making the \emph{Planck} data consistent with the observations of WL, RSD and SZ.

In this work, we wish to discuss how the constraints on the neutrino mass are affected by the parameters of dark energy when the simplest dynamical dark energy models are considered. We do find that the constraints on $\sum m_\nu$ become looser in both the $w$CDM and $w_0w_a$CDM models. But, actually, we should also consider the issue whether more parameters describing the dynamics of dark energy are worthy to be added in the cosmological model in the sense of statistical significance. By simply comparing the minimal $\chi^2$ values of the models in the fits (see Tables~\ref{tab1} and \ref{tab2}; similarly, see also Tables~\ref{tab3} and \ref{tab4} for the cases of considering the inclusion of $N_{\rm eff}$), we find that actually the $\Lambda$CDM cosmology still performs fairly well, since for most cases adding one or two parameters does not improve the fits significantly, i.e. decreases $\chi^2$ by no more than roughly 2, although an exception can also be found (see the Planck+BSH case in Table~\ref{tab1}). For the Planck+BSH case in Table~\ref{tab1}, we see that the $w$CDM model can fit the data best (with $\Delta\chi^2=-3.79$, compared to $\Lambda$CDM), but the $w_0w_a$CDM model does not improve the fit (its $\chi_{\rm min}^2$ is even greatly higher than that of $w$CDM, by $\Delta\chi^2=3.28$). In the whole, we find that neither the $w$CDM model nor the $w_0w_a$CDM model can provide statistically significant improvement over the $\Lambda$CDM model. In particular, the current observations do not seem to favour the $w_0w_a$CDM model that has two more parameters. From the discussion of model selection, we can conclude that the $\Lambda$CDM cosmology can still provide a fairly good description for the current observations and there is no strong support to adding more parameters to describe the dynamics of dark energy.

\subsection{Neutrino mass versus other cosmological parameters in dynamical dark energy models}

Dark energy parameters can affect constraints on neutrino mass, and certainly, it also affects other parameters because of mutual compensation effect in the global fits. To compare the cases of $\Lambda$CDM, $w$CDM and $w_{0}w_{a}$CDM, we use Planck+BAO and Planck+BSH to make an analysis.

\begin{figure*}
\includegraphics[scale=0.45, angle=0]{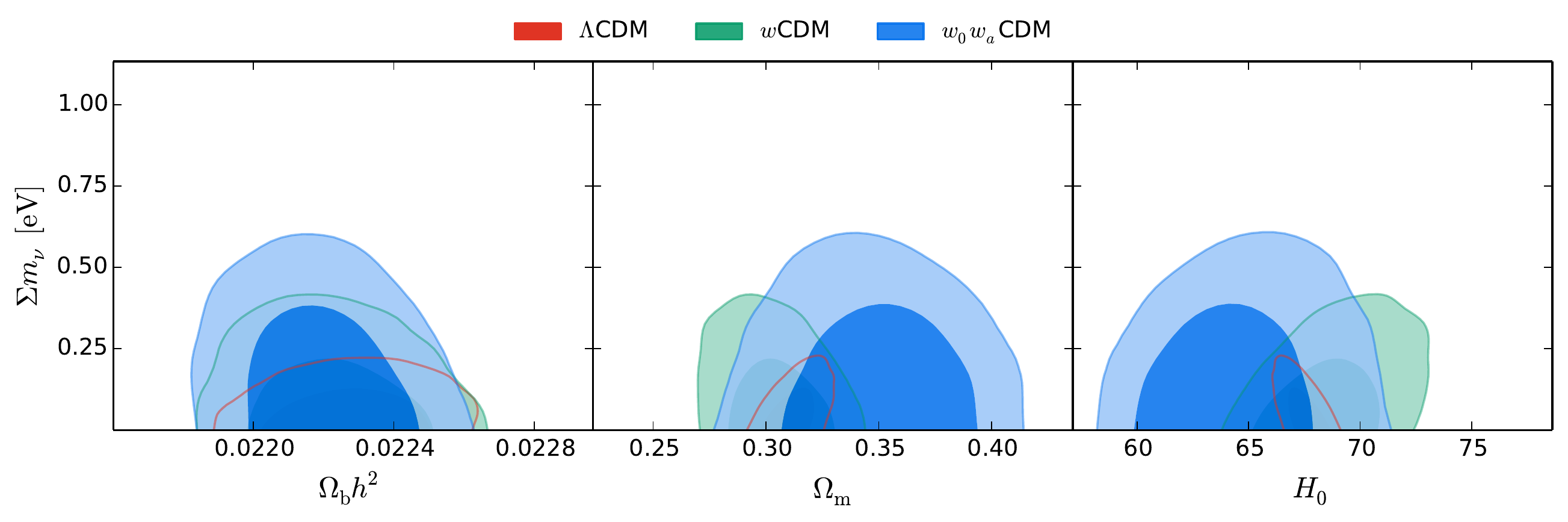}
\includegraphics[scale=0.45, angle=0]{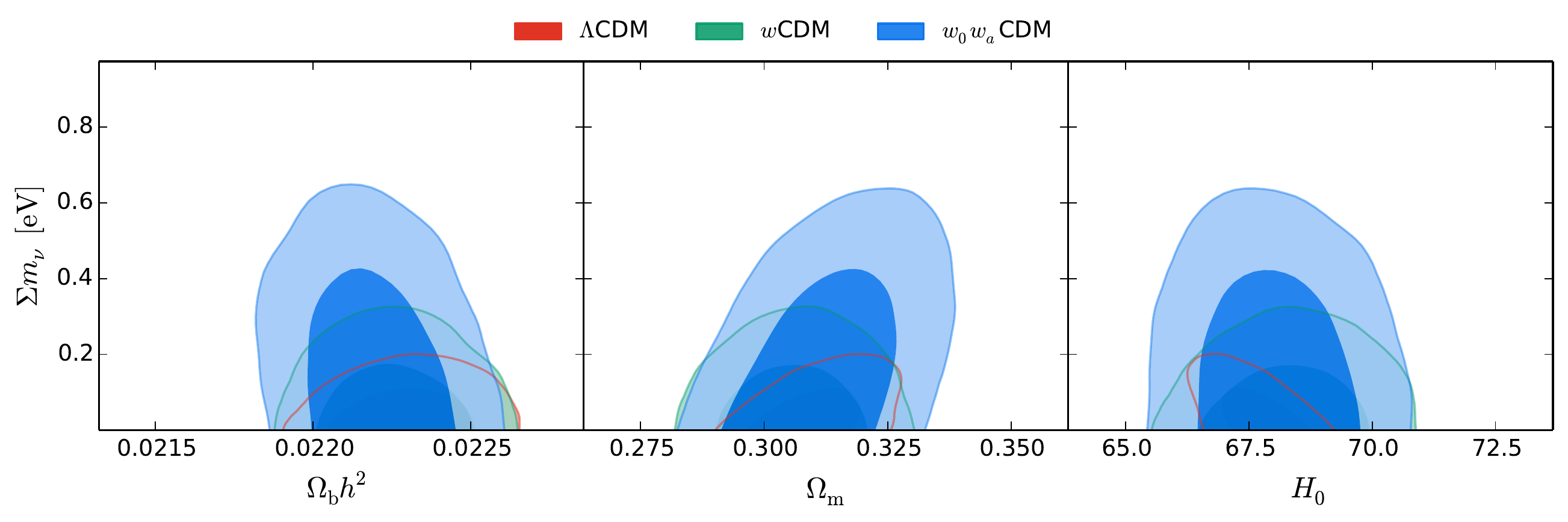}
\caption{The 68 per cent and 95 per cent CL contours in the $\sum m_{\nu}$$-$$\Omega_{\rm b}h^{2}$, $\sum m_{\nu}$$-$$\Omega_{\rm m}$ and $\sum m_{\nu}$$-$$H_{0}$ planes. We show the Planck+BAO constraints in the top panel and the Planck+BSH constraints in the bottom panel.}
\label{fig:late}
\end{figure*}

Fig. \ref{fig:late} shows the constraints on the neutrinos mass and $\Omega_{\rm b}h^{2}$, $\Omega_{\rm m}$ and $H_{0}$ for the three models. We show the constraint results by using the Planck+BAO data combination in the top panel. We find that, the constrained results are quite different for different models. As the contours in the $H_{0}-\sum m_{\nu}$ plane show, $\sum m_{\nu}$ is anti-correlated with $H_{0}$ in $\Lambda$CDM. Here, this correlation in the $\sum m_{\nu}$ and $H_{0}$ can be explained. A larger neutrino mass increases $\theta_{*}$, and a reduction in $H_{0}$ can lead to the same $\theta_{*}$ (a smaller $H_{0}$ corresponds to a larger $D_{A}$ and hence a smaller $\theta_{*}$). However, the correlation direction inverses in the $w$CDM and $w_{0}w_{a}$CDM models, showing a positive correlation between $H_{0}$ and $\sum m_{\nu}$. \emph{Planck} data give a quite precise measurement on $\theta_{*}$, leading to a precise constraint on  $\Omega_{\rm m}h^{3}$ in the $\Lambda$CDM model. But for the $w$CDM and $w_{0}w_{a}$CDM models, $\Omega_{\rm m}h^{3}$ is not constrained well, with a much broader distribution \citep{limiao2013}. However, the distributions of $\Omega_{\rm m}h^{2}$ in the three models are similar. Given the comparison results of $\Omega_{\rm m}h^{2}$ and $\Omega_{\rm m}h^{3}$, we can find a fact that, dynamical dark energy models relax constraints on $H_{0}$. By weaker constraint on $H_{0}$, the tension between \emph{Planck} and the independent measurement of $H_{0}$ can be relieved a little by considering dynamical dark energy \citep{limiao2013}.

The bottom panel of Fig. \ref{fig:late} shows the same case with the Planck+BSH data combination. We find that when the SN and $H_0$ data are combined, all the constraints become tightened, in particular for $\Omega_m$ and $H_0$ in the $w_0w_a$CDM model. As the same to the case of top panel, $\sum m_\nu$ is in anti-correlation with $H_0$ in the $\Lambda$CDM model, while is in slightly positive correlation with $H_0$ in the $w$CDM model and the $w_0w_a$CDM model. Detailed fitting results for all the parameters can be found in Tables \ref{tab1} and \ref{tab2}.

In the $w$CDM model, $\sum m_\nu$ is in anti-correlation with $w$, as explicitly shown in Fig.~\ref{fig:w-mnu}. Also, it is well known that $w$ is in anti-correlation with $H_0$; see, e.g. fig.~21 of \cite{planck2013}. This clearly demonstrates that in the $w$CDM model, $\sum m_\nu$ must be positively correlated with $H_0$. It should also be mentioned that \cite{zhang2016} showed that, besides the $w$CDM model, in the holographic dark energy model, the same conclusion is still kept. Here, in Fig.~\ref{fig:late}, we show that for the $w_{0}w_{a}$CDM model we also have the same conclusion. Therefore, it might be a universal conclusion that in a dynamical dark energy model $\sum m_\nu$ is in positive correlation with $H_0$.

\section{Constraints on the effective number of relativistic species}\label{radiations}

The relativistic energy density in the early universe include the contributions from photons and neutrinos, and possibly other extra relativistic degrees of freedom, called dark radiation. The effective number of relativistic species, including neutrinos and any other dark radiation, is defined by a parameter, $N_{\rm eff}$, for which the standard value is 3.046 corresponding to the case with three-generation neutrinos and no extra dark radiation \citep{Mangano2005}. If the value of $N_{\rm eff}$ is beyond 3.046, it indicates that there is some dark radiation other than three-generation active neutrinos. The behaviour of dark radiation is exactly equivalent to massless neutrinos. Thus, the total radiation energy density in the Universe is given by
\begin{flushright}
\begin{equation}\label{eq:nu}
\rho_{r} = \rho_{\gamma}\left[1+{\frac7 8}\left({\frac4 {11}}\right)^{4 /3} N_{\rm eff}\right],
\end{equation}
\end{flushright}
where $\rho_{\gamma}$ is the energy density of photons. An additional $\Delta N_{\rm eff}$, defined by $N_{\rm eff}-3.046$, if found by observations, indicates the existence of dark radiation, which is important for cosmology. In this section, we will discuss the constraints on $N_{\rm eff}$ in the $w$CDM and $w_0w_a$CDM models.

\subsection{Effects of dark radiation on CMB temperature spectrum}\label{power}

Like massless neutrinos, dark radiation is treated as a free streaming fluid. They do not interact at all for $z\ll10^{10}$. They affect the CMB power spectrum in several ways. First, varying $N_{\rm eff}$ shifts the redshift of matter-radiation equality, $z_{\rm eq}$, defined by
\begin{flushleft}
\begin{equation}\label{equality}
  1+z_{\rm eq}=\frac{\Omega_{\rm m}}{\Omega_{\rm r}}=\frac{\Omega_{\rm m}h^{2}}{\Omega_{\gamma}h^{2}}\frac{1}{1+0.2271N_{\rm eff}}.
\end{equation}
\end{flushleft}
Thus, a larger $N_{\rm eff}$ leads to a reduction in $z_{\rm eq}$, which means the delay of radiation dominance in the Universe, leading to an increase in gravitational potential and an enhancement of the early ISW effect. As a consequence, the first and second peaks of CMB power spectrum are affected.

Secondly, a larger $N_{\rm eff}$ increases the relativistic energy density and hence the expansion rate, which causes an reduction in the comoving sound horizon, $r_{s}$, through $r_{s}\propto 1/H$ \citep{Archidiacono2013}. If $r_{s}$ is decreased, via $\theta_{*}=r_{s}/D_{A}$, $\theta_{*}$ will be decreased as well ($D_{A}$ is less effected in the early Universe). The reduction in $\theta_{*}$ leads the peak positions of the CMB power spectrum to move towards high multipoles.

Third, a larger $N_{\rm eff}$ also enhances the Silk damping tail via expansion rate. The Silk damping is an effect, from the diffusion damping of oscillations in the plasma, caused by an extended decoupling process of baryon$-$photon interactions. The photon freely streams on scale $\lambda_{d}$ within time distance of decoupling, and temperature fluctuation on scale smaller than this scale will be damped. The factor of damping has exp[$-2r_{d}/\lambda_d$], where $r_{d}$ is diffusion length. When diffusion process approaches the last scattering, there is $r_{d}\propto 1/\sqrt{H}$ \citep{Archidiacono2013}, which gives the ratio of $\theta_{d}/\theta_{s}=\sqrt{H}$, where $\theta_{d}$ is damping angular scale. As a consequence, the increased expansion rate by increasing $N_{\rm eff}$ reinforces the Silk damping on small scales.

Lastly, a larger $N_{\rm eff}$ enhances the anisotropic stress at small scales. The distribution function of free streaming species will involve an effect from an extra anisotropic stress when $N_{\rm eff}$ increases, and this stress could change the gravitational potential and hence alter the degree of the reinforced small-scale anisotropy.

\begin{figure}
\begin{center}
\includegraphics[scale=0.3, angle=0]{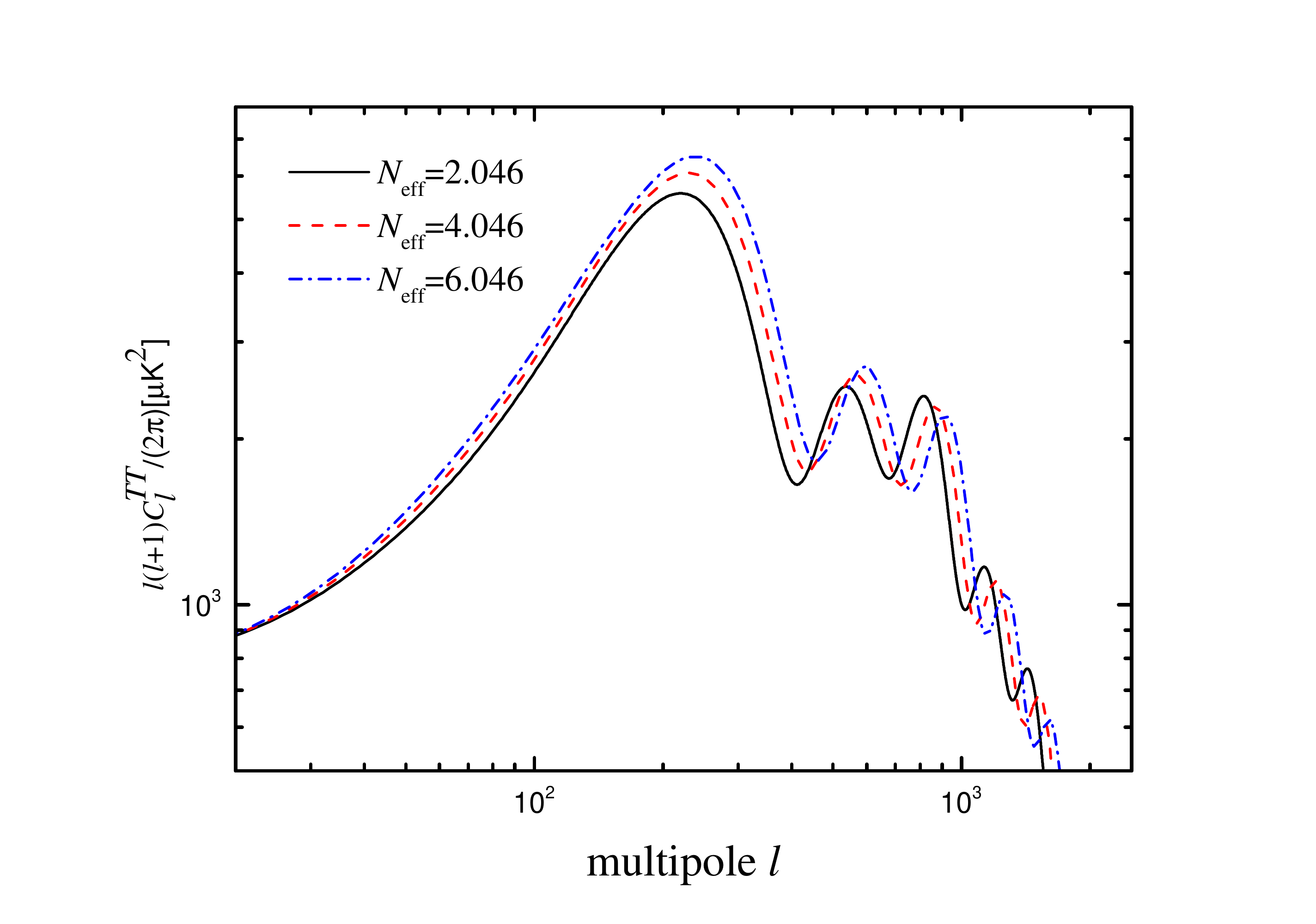}
\end{center}
\caption{The CMB temperature spectra $C_{\ell}^{TT}$ with the different effective numbers of relativistic species $N_{\rm eff}$. We choose $N_{\rm eff}$ $=$ $2.046$, $4.046$ and $6.046$, and fix $w$ $=$ $-1$ and $\sum m_{\nu}$ $=$ $0.06$ eV. At $2<\ell<50$, the temperature power is mildly increased as dark radiation density increases. At $\ell$ $\sim$ $200$, the amplitude of the first peak is enhanced by larger $N_{\rm eff}$ and peak position moves towards high multipoles due to the early ISW effect. On small scales ($\ell$ $>$ $600$), a larger $N_{\rm eff}$ enhances the Silk damping tail of the temperature power.}
\label{fig:HWDEnnu}
\end{figure}

Fig.~\ref{fig:HWDEnnu} shows the $C_{\ell}^{TT}$ spectrum with different $N_{\rm eff}$. We choose the three cases of $N_{\rm eff}$, namely, $N_{\rm eff}=2.046$, 4.046 and 6.046, as examples, and other cosmological parameters are fixed. The figure shows that, at $\ell$ $<$ $600$ multipole, a larger $N_{\rm eff}$ raises the CMB power spectrum. At $\ell<50$ multipole, the power spectrum is mainly affected by the late ISW effect. Around $\ell \sim 200$ scale, the amplitude of peak is largely enhanced, and position of peak moves towards high multipole. At $\ell$ $>$ $600$ scales, we can find that the Silk damping tail is clear. In our analysis, we only concentrate on the effect at large scales of $\ell<200$, related to dark energy.

\subsection{Constraints on dark radiation in dynamical dark energy models}

\begin{figure}
\begin{center}
\includegraphics[scale=1, angle=0]{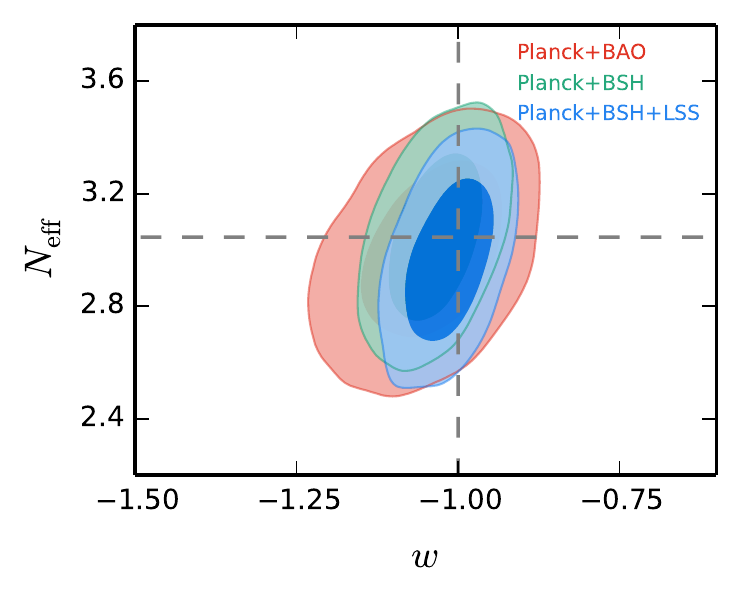}
\end{center}
\caption{68 per cent and 95 per cent CL contours in the $w$$-$$N_{\rm eff}$ plane for the $w$CDM model. The constraints are from the three data combinations, i.e., Planck+BAO, Planck+BSH and Planck+BSH+LSS. Note that the cross point of the grey lines show $w$ $=$ $-1$ and $N_{\rm eff}$ $=$ $3.046$ in the base $\Lambda$CDM. It is shown that the constraints results from the three data combination are all compatible with the base $\Lambda$CDM cosmology.}
\label{fig:wnu}
\end{figure}
\begin{figure}
\begin{center}
\includegraphics[scale=0.4, angle=0]{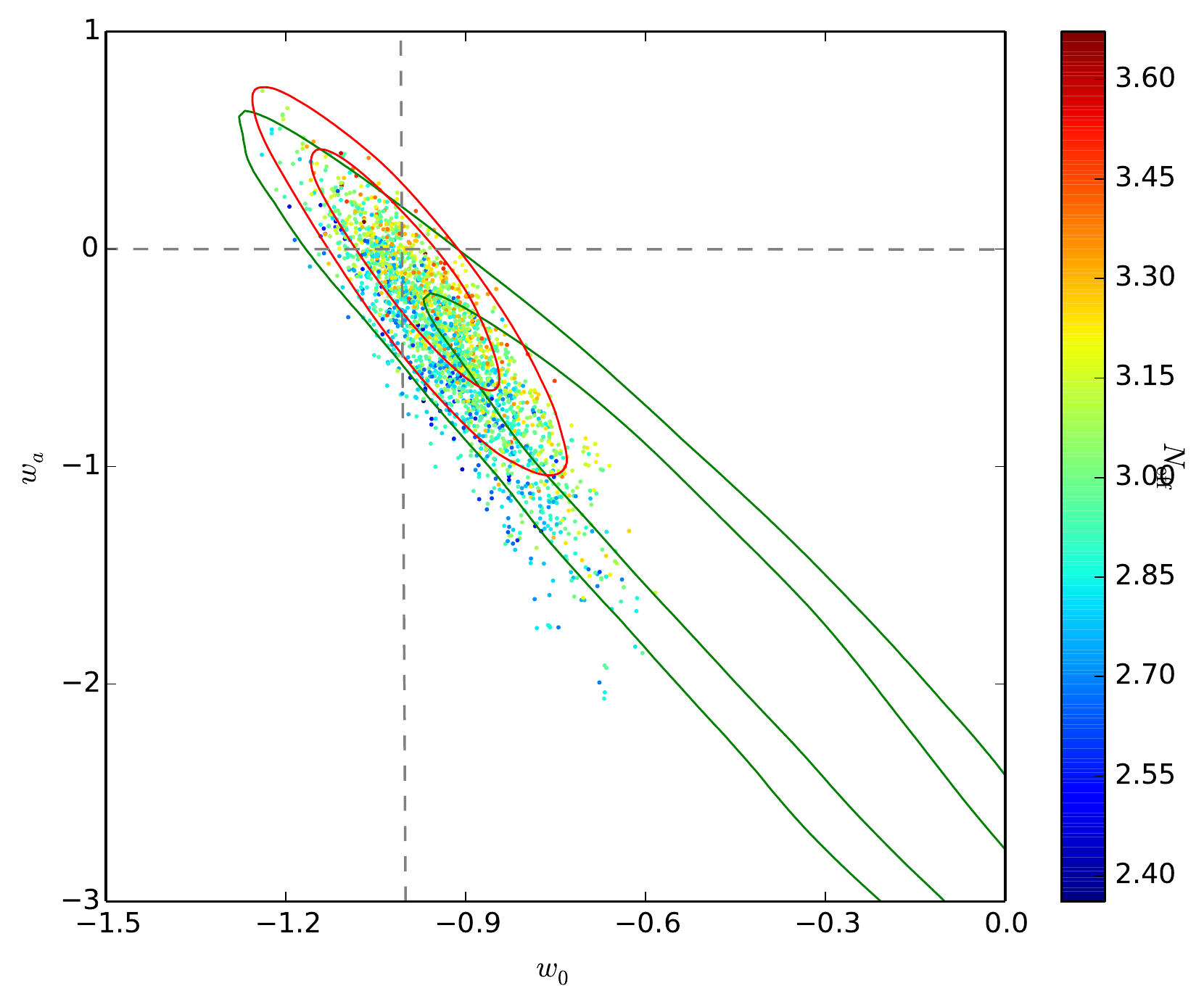}
\caption{Samples from the Planck+BSH chains in the $w_{0}$$-$$w_{a}$ plane, colour-coded by $N_{\rm eff}$. The green contours show the constraints from the Planck+BAO data, and the red contour shows the constraints from the Planck+BSH+LSS combination. The samples show that the points corresponding to smaller $N_{\rm eff}$ distribute the regions of $w$ evolving from $w$ $<$ $-1$ to $w$ $>$ $-1$.}
\label{fig:w0-nnu}
\end{center}
\end{figure}

In this subsection, we study the constraints on dark radiation in $w$CDM and $w_{0}w_{a}$CDM from the Planck+BAO, Planck+BSH and Planck+BSH+LSS data combinations. Depending on the constraint results, we can probe for the correlation between $w$ and $N_{\rm eff}$, and see the effect of dark energy parameter on the cosmological measurement of dark radiation.

The contours in the $N_{\rm eff}$$-$$w$ plane are shown in Fig.~\ref{fig:wnu}. In this figure, we can find that $w$ is slightly positively correlated with $N_{\rm eff}$. This correlation can be explained by the compensation to the effects on the acoustic peak scale $\theta_{*}$. The acoustic peak scale $\theta_{*}$ is determined by $r_{s}/D_{A}$. A larger $w$ leads to a reduction in $D_{A}$ through the increased $H(z)$. If the sound horizon is fixed, $\theta_{*}$ will become larger due to the decreased $D_{A}$. To keep $\theta_{*}$ fixed, the observable $r_{s}$ has to become smaller. Here, increasing $N_{\rm eff}$ can lead to a smaller sound horizon $r_{s}$ through the increased $H(z)$. By comparison to the constraint results of $N_{\rm eff}$ from the three data combinations, we find that the Planck+BSH data combination gives the largest value of $N_{\rm eff}$. Once the LSS observations are included, the value of $N_{\rm eff}$ becomes smaller. However, the constraint results of $N_{\rm eff}$ from the three data combinations are all compatible with the standard value of $3.046$, which means that there is no evidence of deviation from the standard model of particle physics. Also, from Fig.~\ref{fig:wnu}, we see that the cosmological constant $\Lambda$ ($w=-1$) is consistent with the current data.

For the case of $w_{0}w_{a}$CDM model, we plot the marginalized posterior contours in the $w_{0}$$-$$w_{a}$ plane from the Planck+BAO and Planck+BSH+LSS data combinations, and present them in Fig.~\ref{fig:w0-nnu}. The samples from Planck+BSH chains in the $w_{0}$$-$$w_{a}$ plane are also shown in Fig.~\ref{fig:w0-nnu}, colour coded by the value of $N_{\rm eff}$. In this figure, we can find that a smaller $N_{\rm eff}$ is allowed by an early phantom dark energy evolving from $w<-1$ to $w>-1$. As same as the $w_{0}$CDM model, the $w_{0}w_{a}$CDM model also allows for a standard value of $N_{\rm eff}$ from the three data combinations.

Next, we compare the constraint results of $N_{\rm eff}$ for the $\Lambda$CDM, $w$CDM and $w_{0}w_{a}$CDM models. The fitting results are displayed in Tables~\ref{tab3} and \ref{tab4}.

The Planck+BAO data combination gives the constraints: $N_{\rm eff}=3.04^{+0.19}_{-0.18}$ for $\Lambda$CDM, $N_{\rm eff}=3.00\pm0.20$ for $w$CDM and $N_{\rm eff}=2.96\pm0.20$ for $w_{0}w_{a}$CDM. Correspondingly, we have $w=-1.042^{+0.076}_{-0.065}$ for $w$CDM, $w_{0}=-0.50^{+0.36}_{-0.25}$ and $w_{a}=-1.53^{+0.73}_{-1.08}$ for $w_{0}w_{a}$CDM. From these results, we can find that the $w$CDM and $w_{0}w_{a}$CDM models allow a smaller $N_{\rm eff}$ compared to $\Lambda$CDM. For the case of dynamical dark energy, we can find that the $w$CDM and $w_{0}w_{a}$CDM models are in favour of a phantom energy with $w<-1$ and a quintom energy evolving from $w<-1$ to $w>-1$, respectively. Therefore, a smaller $N_{\rm eff}$ is favoured by the two dynamical dark energy models much better than the $\Lambda$CDM model due to the positive correlation between $N_{\rm eff}$ and $w$.

In the joint fits to Planck+BSH data, we have $N_{\rm eff}=3.11\pm0.17$ for $\Lambda$CDM, $N_{\rm eff}=3.05^{+0.18}_{-0.19}$ for $w$CDM and $N_{\rm eff}=2.99^{+0.21}_{-0.19}$ for $w_{0}w_{a}$CDM. Correspondingly, we have $w=-1.034\pm0.045$ for $w$CDM and $w_{0}=-0.92^{+0.09}_{-0.12}$ and $w_{a}=-0.45^{+0.49}_{-0.35}$ for $w_{0}w_{a}$CDM. From these results, we clearly find that the constraints on the EoS of dark energy are tightened by the inclusion of SN and $H_{0}$ data. However, by adding the SN and $H_{0}$ data, the constraints of dark radiation are looser in three models. For the $\Lambda$CDM model, the Planck+BSH data combination gives obviously looser constraint on $N_{\rm eff}$ than the Planck+BAO data combination. For the $w$CDM model and the $w_{0}w_{a}$CDM model, the Planck+BSH data give slightly looser constraints on $N_{\rm eff}$ than the Planck+BAO data.

Further considering the LSS observations in the combination, the constraint results of dark radiation become $N_{\rm eff}=2.99\pm 0.16$ for $\Lambda$CDM, $N_{\rm eff}=2.97^{+0.18}_{-0.19}$ for $w$CDM and $N_{\rm eff}=2.95\pm0.19$ for $w_{0}w_{a}$CDM, respectively. For the EoS of dark energy, we have $w=-1.014\pm0.042$ for $w$CDM and $w_{0}=-0.99\pm0.09$ and $w_{a}=-0.09^{+0.41}_{-0.32}$ for $w_{0}w_{a}$CDM. We can find that the Planck+BSH+LSS data combination has little contribution to the constraints on dark energy, but gives the tightest constraints on dark radiation compared to the Planck+BAO and Planck+BSH data combinations. For dark energy, as mentioned above, the LSS observations can not provide the tight constraints on $w$ due to the effects of dark energy on the growth of structure only through the expansion history. For $N_{\rm eff}$, the LSS observations prefer a lower $\sigma_{8}$, which also leads to smaller values of $N_{\rm eff}$ due to the positive correlation between $\sigma_{8}$ and $N_{\rm eff}$.

\begin{figure*}
\begin{center}
\includegraphics[scale=0.45, angle=0]{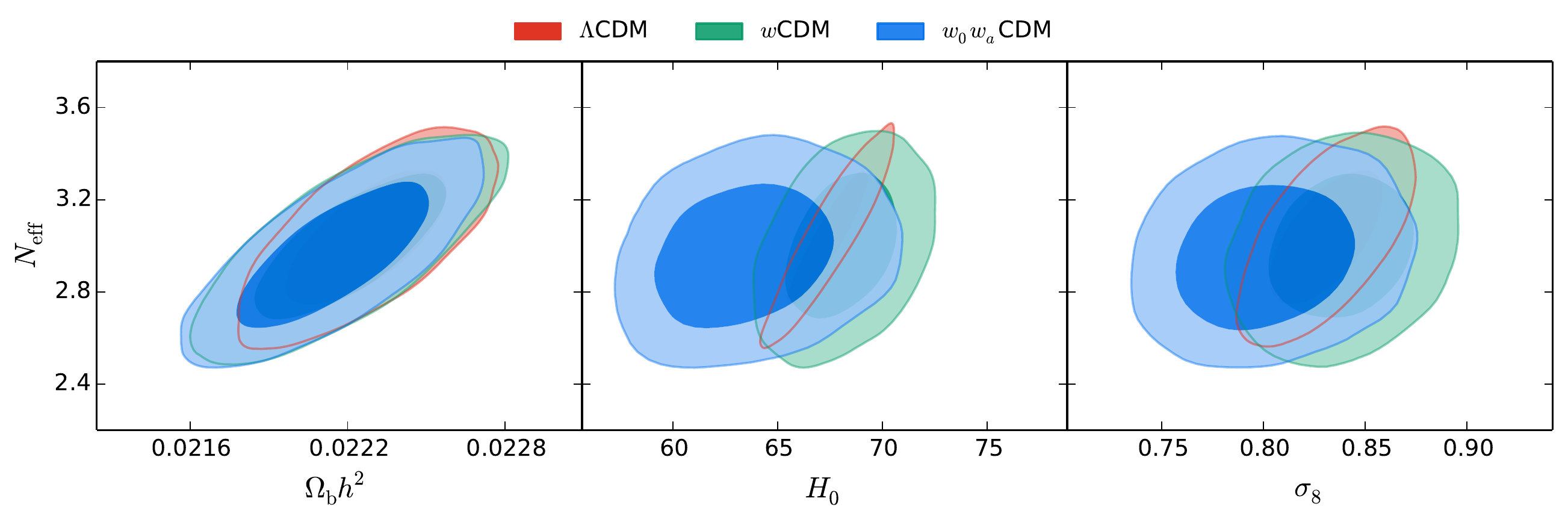}
\includegraphics[scale=0.45, angle=0]{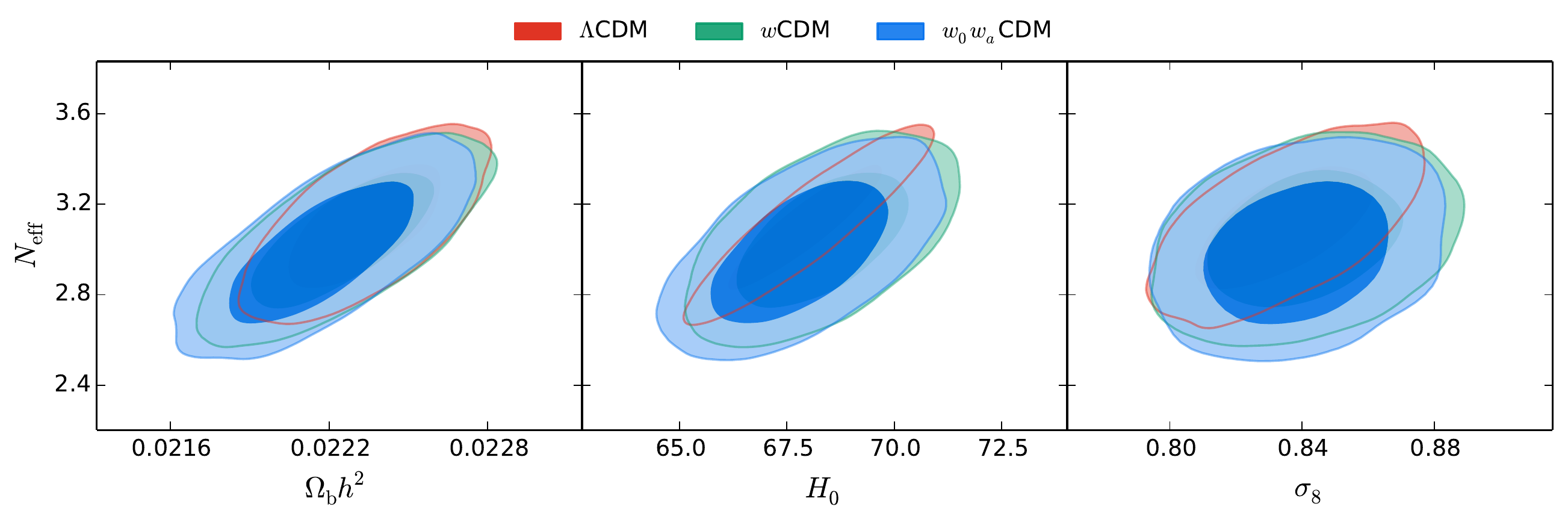}
\caption{The 68 per cent and 95 per cent CL contours in the $N_{\rm eff}$$-$$\Omega_{\rm b}h^{2}$, $N_{\rm eff}$$-$$H_0$ and $N_{\rm eff}$$-$$\sigma_{8}$ planes. We show the Planck+BAO constraints in the top panel and the Planck+BSH constraints in the bottom panel.}
\label{fig:lateparameter}
\end{center}
\end{figure*}

Fig.~\ref{fig:lateparameter} shows the constraints on $N_{\rm eff}$ and $\Omega_{\rm b}h^{2}$, $H_{0}$ and $\sigma_{8}$ for three models. The Planck+BAO data give the $\Omega_{\rm b}h^{2}$$-$$N_{\rm eff}$, $\sigma_{8}$$-$$N_{\rm eff}$, and $H_{0}$$-$$N_{\rm eff}$ contours in the top panel and the Planck+BSH data give the corresponding contours in the bottom panel. From the figure, we can find that the Planck+BAO data give the consistent constraint contours in the $\Omega_{\rm b}h^{2}$$-$$N_{\rm eff}$, but a little bit different constraint contours in the $H_{0}$$-$$N_{\rm eff}$ and $\sigma_{8}$$-$$N_{\rm eff}$ planes for different models. Here, we focus on the $H_{0}$$-$$N_{\rm eff}$ plane. As the constraint contours show, the correlation between $N_{\rm eff}$ and $H_{0}$ is consistent in the $\Lambda$CDM, $w$CDM and $w_{0}w_{a}$CDM models, i.e. $N_{\rm eff}$ is positively correlated with $H_{0}$. This correlation can be illustrated. \emph{Planck} accurately measures the acoustic scale $r_{*}/D_{A}$. Increasing $N_{\rm eff}$ leads the sound horizon at recombination to be smaller, and hence recombination has to be closer (larger $H_{0}$ and hence smaller $D_{A}$) for it to keep the same angular size observed by \emph{Planck} \citep{planck2015a}. Therefore, a larger $N_{\rm eff}$ favours a higher $H_{0}$. The tension between \emph{Planck} and the direct measurement of Hubble constant can be relieved by considering dark radiation in the cosmological models. However, for the constraint values of $H_{0}$, the $w$CDM model favours a relatively larger value of $H_{0}$, having $H_{0}=68.1\pm1.7$ $\rm km \, s^{-1}Mpc^{-1}$, and the $w_{0}w_{a}$CDM model favours a lower value of $H_{0}$, having $H_{0}=63.6^{+2.3}_{-3.2}$ $\rm km \, s^{-1}Mpc^{-1}$. This is because the $w$CDM model and the $w_{0}w_{a}$CDM model relax the constraints on $H_{0}$ due to the weaker constraints on dark energy under Planck+BAO, especially for the $w_{0}w_{a}$CDM model.

The bottom panel of Fig.~\ref{fig:lateparameter} shows the same constraints cases from the Planck+BSH data combination. We find that once the SN and $H_{0}$ data are included, all parameter spaces are shrunk, in particular for the parameter $H_{0}$ in the $w_{0}w_{a}$CDM model.

In summary, the current observations favour the standard result of $N_{\rm eff}=3.046$ and $w=-1$ for all the three models. This also indicates that the dark energy parameters actually have no impact on the constraint of $N_{\rm eff}$.

\section{Conclusion}\label{sec:conclusion}

\begin{figure*}
\begin{center}
\includegraphics[scale=1.05, angle=0]{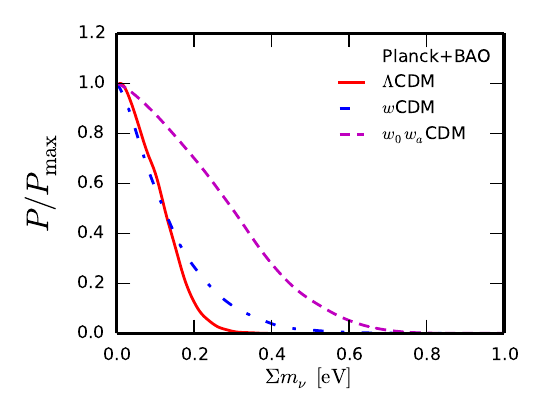}
\includegraphics[scale=1.05, angle=0]{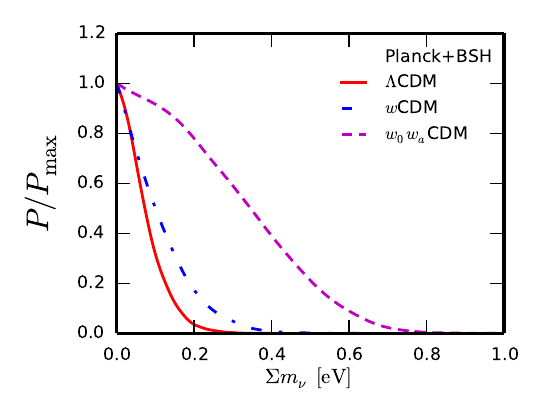}
\includegraphics[scale=1.05, angle=0]{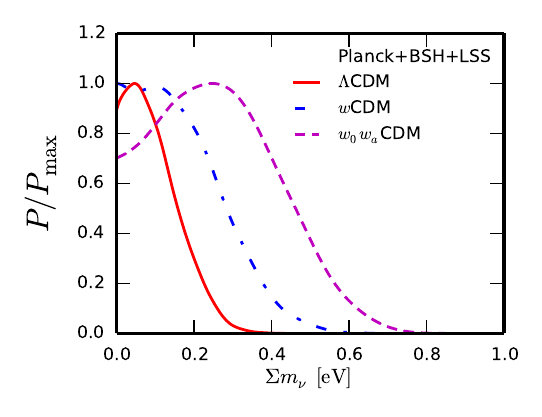}
\caption{The one-dimensional marginalized distributions of $\sum m_{\nu}$ for $\Lambda$CDM (red solid), $w$CDM (blue dash$-$dotted) and $w_{0}w_{a}$CDM (purple dashed) under the constraints of Planck+BAO, Planck+BSH and Planck+BSH+LSS, respectively.}
\label{fig:likelihood1}
\end{center}
\end{figure*}

\begin{figure*}
\begin{center}
\includegraphics[scale=0.6, angle=0]{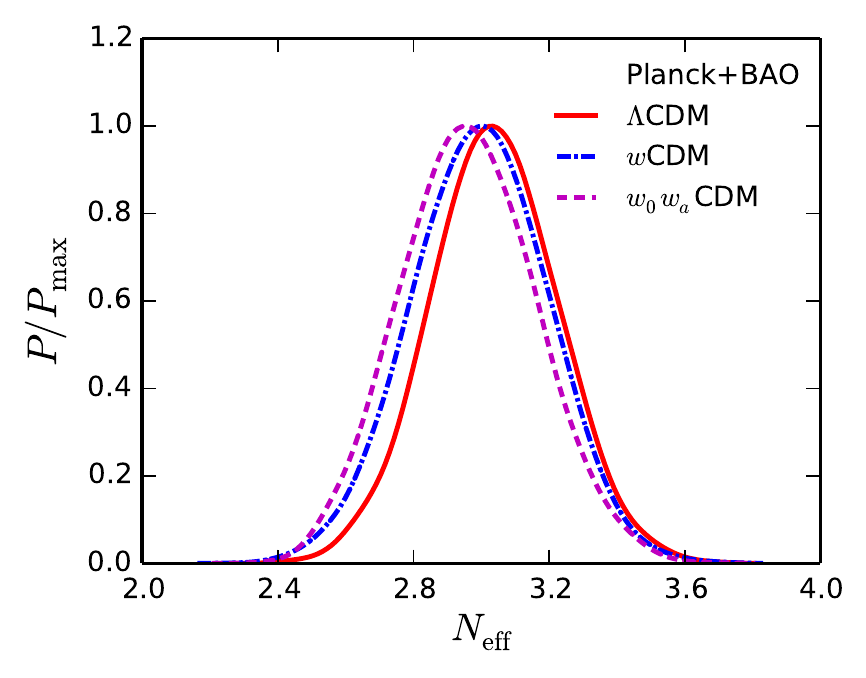}
\includegraphics[scale=0.6, angle=0]{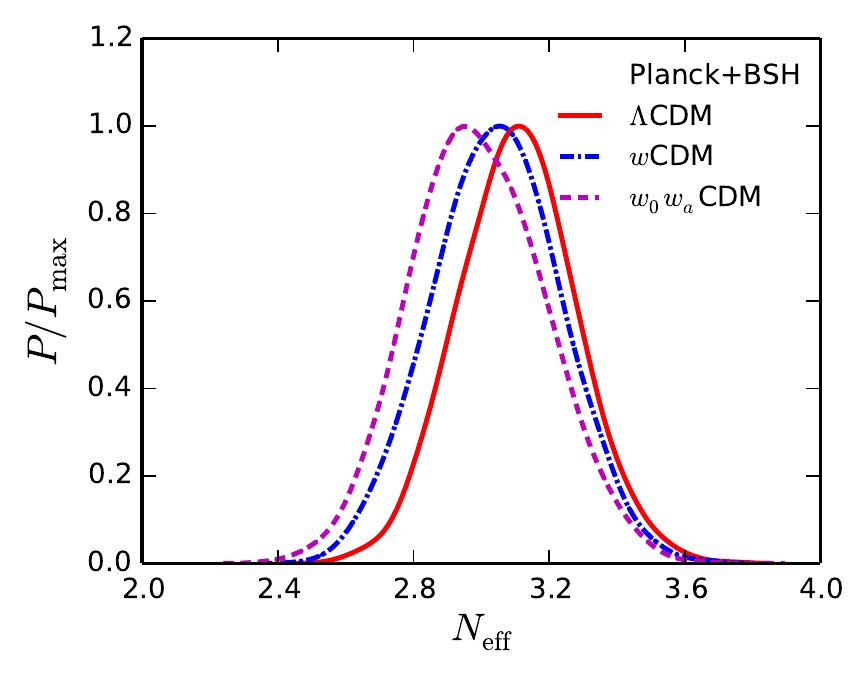}
\includegraphics[scale=0.6, angle=0]{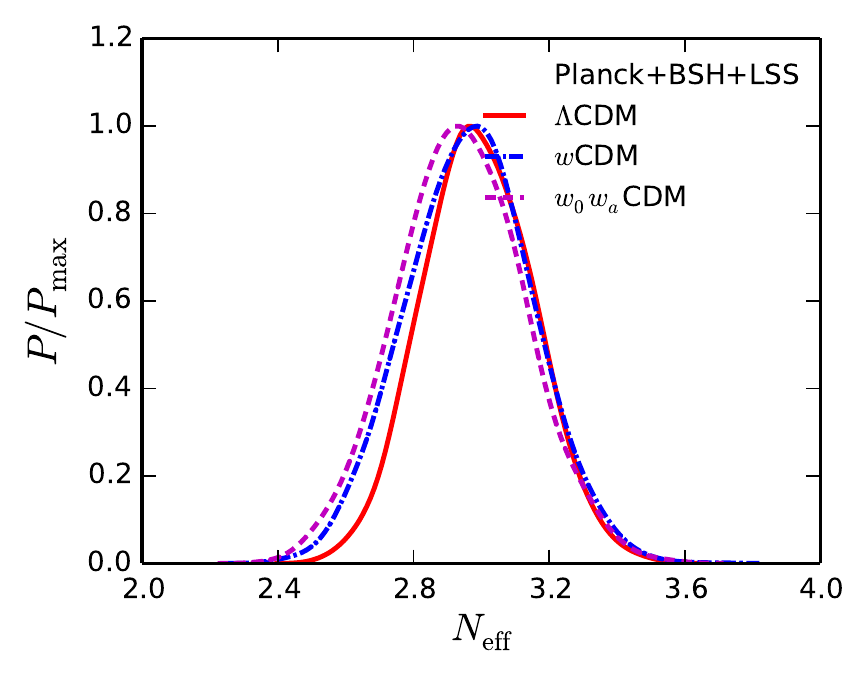}
\caption{The one-dimensional marginalized distributions of $N_{\rm eff}$ for $\Lambda$CDM (red solid), $w$CDM (blue dash$-$dotted), and $w_{0}w_{a}$CDM (purple dashed) under the constraints of Planck+BAO, Planck+BSH and Planck+BSH+LSS, respectively.}
\label{fig:likelihood2}
\end{center}
\end{figure*}

In this paper, we investigate how the dark energy parameters affect the cosmological constraints on the neutrino mass $\sum m_\nu$ and the effective number of relativistic species $N_{\rm eff}$. We only consider the most basic extensions of the $\Lambda$CDM cosmology, i.e. the $w$CDM model and the $w_0w_a$CDM model. We use the latest cosmological observations to constrain the neutrino mass and the extra relativistic degrees of freedom in these models and make comparison for them. We choose three data combinations to do the global fits, which are Planck+BAO, Planck+BSH and Planck+BSH+LSS. We wish to give a uniform comparison of the constraints on $\sum m_\nu$ and $N_{\rm eff}$ in $\Lambda$CDM, $w$CDM and $w_0w_a$CDM under the same conditions. Note that we separately constrain $\sum m_\nu$ and $N_{\rm eff}$ in these models.

We give the 95 per cent CL upper limits of $\sum m_\nu$. (i) Using Planck+BAO, we obtain $\sum m_{\nu}<0.17$ eV for the $\Lambda$CDM model, $\sum m_{\nu}<0.33$ eV for the $w$CDM model and $\sum m_{\nu}<0.47$ eV for the $w_{0}w_{a}$CDM model. (ii) Using Planck+BSH, we obtain $\sum m_{\nu}<0.15$ eV for the $\Lambda$CDM model, $\sum m_{\nu}<0.25$ eV for the $w$CDM model and $\sum m_{\nu}<0.51$ eV for the $w_{0}w_{a}$CDM model. (iii) Using Planck+BSH+LSS, we obtain $\sum m_{\nu}<0.22$ eV for the $\Lambda$CDM model, $\sum m_{\nu}<0.36$ eV for the $w$CDM model and $\sum m_{\nu}<0.52$ eV for the $w_{0}w_{a}$CDM model.

The comparison of these results is briefly summarized in Fig.~\ref{fig:likelihood1}, which shows the one-dimensional posterior distributions of $\sum m_\nu$ in the three models using the Planck+BAO, Planck+BSH and Planck+BSH+LSS data combinations, respectively. We find that the dynamical dark energy models, both $w$CDM and $w_0w_a$CDM, allow for a larger upper limit of $\sum m_\nu$. Though the cosmological constant $\Lambda$ is still consistent with the current data, our analysis shows that a phantom dark energy ($w<-1$) or an early phantom dark energy (i.e. quintom evolving from $w<-1$ to $w>-1$) is slightly more favoured by current observations. This leads to the fact that in both $w$CDM and $w_0w_a$CDM we obtain a larger upper limit of $\sum m_\nu$. The correlation between dark energy parameter and neutrino mass is discussed in detail and in depth in this paper.

Furthermore, we give the constraint results of $N_{\rm eff}$. (i) Using Planck+BAO, we obtain $N_{\rm eff}=3.14^{+0.19}_{-0.18}$ for the $\Lambda$CDM model, $N_{\rm eff}=3.00\pm0.20$ for the $w$CDM model and $N_{\rm eff}=2.96\pm0.20$ for the $w_{0}w_{a}$CDM model. (ii) Using Planck+BSH, we obtain $N_{\rm eff}=3.11\pm0.17$ for the $\Lambda$CDM model, $N_{\rm eff}=3.05^{+0.18}_{-0.21}$ for the $w$CDM model and $N_{\rm eff}=2.99^{+0.19}_{-0.21}$ for the $w_{0}w_{a}$CDM model. (iii) Using Planck+BSH+LSS, we obtain $N_{\rm eff}=2.99\pm 0.16$ for the $\Lambda$CDM model, $N_{\rm eff}=2.97^{+0.18}_{-0.19}$ for the $w$CDM model and $N_{\rm eff}=2.95\pm0.19$ for the $w_{0}w_{a}$CDM model.

The comparison of these results is briefly summarized in Fig.~\ref{fig:likelihood2}, which shows the one-dimensional posterior distributions of $N_{\rm eff}$ in the three models using the Planck+BAO, Planck+BSH and Planck+BSH+LSS data combinations, respectively. We clearly show that in the three dark energy models the constraints on $N_{\rm eff}$ are in good accordance with each other, all in favour of the standard value 3.046. This indicates that the dark energy parameters almost have no impact on constraining $N_{\rm eff}$.

Therefore, we clearly show that the dark energy parameters can exert a significant influence on the cosmological weighing of neutrinos, but almost cannot affect the constraint on the extra relativistic degrees of freedom.

\begin{table*}
\caption{\label{tab1} Fitting results for the $\Lambda$CDM, $w$CDM and $w_{0}w_{a}$CDM models from the Planck+BAO and Planck+BSH data combinations, respectively. Here, we fix $N_{\rm eff}$$=$$3.046$. We quote the $\pm 1\sigma$ errors, but for the neutrino mass $\sum m_{\nu}$, we quote the 95 per cent CL upper limits. Note that $\sum m_{\nu}$ is in units of eV, and $H_{0}$ is in units of $\rm km\;s^{-1} Mpc^{-1}$.}
\centering
\begin{tabular}{ccccccccc}
\hline  \multicolumn{1}{c}{Data}&&\multicolumn{3}{c}{Planck+BAO}&&\multicolumn{3}{c}{Planck+BSH}\\
          \cline{1-1}\cline{3-5}\cline{7-9}

       Model&&$\Lambda$CDM & $w$CDM &$w_{0}w_{a}$CDM&&$\Lambda$CDM&$w$CDM&$w_{0}w_{a}$CDM\\
\hline

$\Omega_{\rm b}h^2$&&$0.02228\pm0.00015$&$0.02223^{+0.00016}_{-0.00015}$&$0.02220\pm0.00015$&&$0.02230\pm0.00014$&$0.02226\pm0.00015$&$0.02219\pm0.00015$\\
$\Omega_{\rm c}h^2$&&$0.1192\pm0.0011$&$0.1197\pm0.0014$&$0.1200\pm0.0014$&&$0.1191\pm0.0011$&$0.1195\pm0.0013$&$0.1201^{+0.0014}_{-0.0013}$\\
$100\theta_{\rm MC}$&&$1.04083^{+0.0003}_{-0.00031}$&$1.04075^{+0.00032}_{-0.00031}$&$1.04069\pm0.00032$&&$1.04086\pm0.00030$&$1.04078\pm0.00031$&$1.04068\pm0.00031$\\
$\tau$&&$0.082\pm0.017$&$0.081\pm0.018$&$0.079^{+0.018}_{-0.017}$&&$0.083\pm0.017$&$0.081\pm0.017$&$0.080^{+0.017}_{-0.018}$\\
$w/w_{0}$&&$-$&$-1.068^{+0.103}_{-0.067}$&$-0.52^{+0.34}_{-0.22}$&&$-$&$-1.042^{+0.052}_{-0.045}$&$-0.89^{+0.12}_{-0.14}$\\
$w_a$&&$-$&$-$&$-1.73^{+0.39}_{-1.25}$&&$-$&$-$&$-0.84^{+0.80}_{-0.49}$\\
$\Sigma m_\nu$&&$<0.17$&$<0.33$&$<0.47$&&$<0.15$&$<0.25$&$<0.51$\\
$n_s$&&$0.9659\pm0.0041$&$0.9645\pm0.0046$&$0.9630^{+0.0046}_{-0.0045}$&&$0.9664\pm0.0041$&$0.9652\pm0.0044$&$0.9626^{+0.0046}_{-0.0047}$\\
${\rm{ln}}(10^{10}A_s)$&&$3.097\pm0.033$&$3.096^{+0.035}_{-0.034}$&$3.093\pm0.034$&&$3.099\pm0.033$&$3.096^{+0.033}_{-0.032}$&$3.095\pm0.034$\\
$\Omega_{\rm m}$&&$0.3128^{+0.0073}_{-0.0075}$&$0.3040\pm0.0140$&$0.3490^{+0.0300}_{-0.0250}$&&$0.3109^{+0.0067}_{-0.0075}$&$0.3060\pm0.0090$&$0.3130\pm0.0110$\\
$H_0$&&$67.4^{+0.6}_{-0.5}$&$68.7^{+1.6}_{-1.9}$&$64.5^{+2.2}_{-3.1}$&&$67.6^{+0.6}_{-0.5}$&$68.31\pm1.0$&$68.0\pm1.1$\\
$\sigma_8$&&$0.829^{+0.019}_{-0.016}$&$0.836\pm0.023$&$0.792^{+0.029}_{-0.033}$&&$0.831^{+0.018}_{-0.015}$&$0.835^{+0.020}_{-0.019}$&$0.821^{+0.031}_{-0.024}$\\

\hline
$\chi^{2}_{\rm min}$ &&12 940.94 &12 939.28 & 12 938.46 && 13 657.29&13 653.50&13 656.78\\
\hline
\end{tabular}
\end{table*}

\begin{table*}
\caption{\label{tab2} Fitting results for the $\Lambda$CDM, $w$CDM and $w_{0}w_{a}$CDM from the Planck+BSH+LSS data combination. Here, we fix $N_{\rm eff}$$=$$3.046$. We quote the $\pm 1\sigma$ errors, but for the neutrino mass $\sum m_{\nu}$, we quote the 95 per cent CL upper limits. Note that $\sum m_{\nu}$ is in units of eV, and $H_{0}$ is in units of $\rm km\;s^{-1} Mpc^{-1}$.}
\centering
\begin{tabular}{ccccccccc}
\hline \multicolumn{1}{c}{Data} &&\multicolumn{3}{c}{Planck+BSH+LSS}&\\
      \cline{1-1}\cline{3-5}

 Model &&$\Lambda$CDM& $w$CDM&$w_{0}w_{a}$CDM\\

\hline
$\Omega_{\rm b}h^2$&&$0.02235\pm0.00014$&$0.02231\pm0.00014$&$0.02227\pm0.00015$\\
$\Omega_{\rm c}h^2$&&$0.1178\pm0.0011$&$0.1181\pm0.0012$&$0.1184\pm0.0012$\\
$100\theta_{\rm MC}$&&$1.04096\pm0.00030$&$1.04090\pm0.00030$&$1.04083\pm0.00031$\\
$\tau$&&$0.066^{+0.014}_{-0.016}$&$0.067\pm0.015$&$0.068\pm0.015$\\
$w/w_{0}$&&$-$&$-1.042^{+0.057}_{-0.047}$&$-0.96\pm0.11$\\
$w_a$&&$-$&$-$&$-0.47^{+0.59}_{-0.43}$\\
$\Sigma m_\nu$&&$<0.22$&$<0.36$&$<0.52$\\
$n_s$&&$0.9684^{+0.0040}_{-0.0041}$&$0.9672^{+0.0042}_{-0.0043}$&$0.9656\pm0.0046$\\
${\rm{ln}}(10^{10}A_s)$&&$3.062^{+0.027}_{-0.030}$&$3.064\pm0.029$&$3.065\pm0.028$\\
$\Omega_m$&&$0.3072^{+0.0071}_{-0.0082}$&$0.3053^{+0.0085}_{-0.0086}$&$0.3090^{+0.0100}_{-0.0101}$\\
$H_0$&&$67.8^{+0.7}_{-0.6}$&$68.3\pm1.0$&$68.2\pm1.0$\\
$\sigma_8$&&$0.803^{+0.015}_{-0.012}$&$0.800^{+0.017}_{-0.014}$&$0.791^{+0.022}_{-0.019}$\\

\hline
$\chi^{2}_{\rm min}$ &&13 906.47 &13 905.66 & 13 904.06\\
\hline
\end{tabular}
\label{tab:overallneff}
\end{table*}

\begin{table*}
\caption{\label{tab3} Fitting results for the $\Lambda$CDM, $w$CDM and $w_{0}w_{a}$CDM models from the Planck+BAO and Planck+BSH data combinations, respectively. Here, we fix $\sum m_{\nu}$$=$$0.06$ eV. We quote the $\pm 1\sigma$ errors. Note that $H_{0}$ is in units of $\rm km\;s^{-1} Mpc^{-1}$.}
\centering
\begin{tabular}{ccccccccc}
\hline  \multicolumn{1}{c}{Data}&&\multicolumn{3}{c}{Planck+BAO}&&\multicolumn{3}{c}{Planck+BSH}\\
          \cline{1-1}\cline{3-5}\cline{7-9}

        Model&&$\Lambda$CDM & $w$CDM &$w_{0}w_{a}$CDM&&$\Lambda$CDM&$w$CDM&$w_{0}w_{a}$CDM\\
\hline

$\Omega_{\rm b}h^2$&&$0.02228\pm0.00019$&$0.02221\pm0.00023$&$0.02215\pm0.00023$&&$0.02234\pm0.00019$&$0.02225^{+0.00022}_{-0.00021}$&$0.02217^{+0.00022}_{-0.00024}$\\
$\Omega_{\rm c}h^2$&&$0.1192^{+0.0031}_{-0.0033}$&$0.1191^{+0.0030}_{-0.0031}$&$0.1189\pm0.0031$&&$0.1200^{+0.0029}_{-0.0030}$&$0.1196\pm0.0030$&$0.1194^{+0.0031}_{-0.0030}$\\
$100\theta_{\rm MC}$&&$1.04085^{+0.00044}_{-0.00045}$&$1.04087^{+0.00044}_{-0.00043}$&$1.04088^{+0.00044}_{-0.00047}$&&$1.04076\pm0.00042$&$1.0408^{+0.00042}_{-0.00046}$
&$1.04081^{+0.00045}_{-0.00044}$\\
$\tau$&&$0.082^{+0.016}_{-0.017}$&$0.078^{+0.019}_{-0.018}$&$0.074\pm0.018$&&$0.084\pm0.016$&$0.080\pm0.017$&$0.073^{+0.017}_{-0.016}$\\
$w/w_{0}$&&$-$&$-1.042^{+0.076}_{-0.065}$&$-0.50^{+0.36}_{-0.25}$&&$-$&$-1.034\pm0.045$&$-0.92^{+0.09}_{-0.11}$\\
$w_a$&&$-$&$-$&$-1.53^{+0.73}_{-1.08}$&&$-$&$-$&$-0.45^{+0.49}_{-0.35}$\\
$N_{\rm eff}$&&$3.04^{+0.19}_{-0.18}$&$3.00\pm0.20$&$2.96\pm0.20$&&$3.11\pm0.17$&$3.05^{+0.18}_{-0.19}$&$2.99^{+0.19}_{-0.21}$\\
$n_s$&&$0.9657\pm0.0075$&$0.9627\pm0.0091$&$0.9605\pm0.009$&&$0.9686\pm0.0069$&$0.965^{+0.0083}_{-0.0084}$&$0.9611\pm0.009$\\
${\rm{ln}}(10^{10}A_s)$&&$3.097^{+0.035}_{-0.036}$&$3.088^{+0.040}_{-0.039}$&$3.080^{+0.037}_{-0.040}$&&$3.103\pm0.034$&$3.095^{+0.039}_{-0.036}$&$3.080^{+0.035}_{-0.034}$\\
$\Omega_m$&&$0.3125\pm0.0075$&$0.3060\pm0.0130$&$0.3520^{+0.0320}_{-0.0270}$&&$0.3096^{+0.0070}_{-0.0069}$&$0.3054^{+0.0089}_{-0.0090}$&$0.3094^{+0.0097}_{-0.0100}$\\
$H_0$&&$67.4\pm1.2$&$68.1\pm1.7$&$63.6^{+2.3}_{-3.2}$&&$68.0\pm1.1$&$68.3\pm1.2$&$67.8^{+1.3}_{-1.4}$\\
$\sigma_8$&&$0.831\pm0.017$&$0.839^{+0.022}_{-0.023}$&$0.802^{+0.025}_{-0.031}$&&$0 .835\pm0.017$&$0 .841\pm0.018$&$0 .839^{+0.018}_{-0.017}$\\

\hline
$\chi^{2}_{\rm min}$ &&12 950.28 &12 950.23 & 12 947.26 && 13 657.29&13 656.34&13 655.36 \\
\hline
\end{tabular}
\label{tab:overall2}
\end{table*}

\begin{table*}
\caption{\label{tab4}Fitting results for the $\Lambda$CDM, $w$CDM and $w_{0}w_{a}$CDM models from the Planck+BSH+LSS data combination. Here, we fix $\sum m_{\nu}$$=$$0.06$ eV. We quote the $\pm 1\sigma$ errors. Note $H_{0}$ is in units of $\rm km\;s^{-1} Mpc^{-1}$.}
\centering
\begin{tabular}{ccccccccc}
\hline \multicolumn{1}{c}{Data} &&\multicolumn{3}{c}{Planck+BSH+LSS}&\\
      \cline{1-1}\cline{3-5}

 Model &&$\Lambda$CDM& $w$CDM&$w_{0}w_{a}$CDM\\

\hline
$\Omega_{\rm b}h^2$&&$0.02231\pm0.00018$&$0.02227^{+0.00021}_{-0.00023}$&$0.02225\pm0.00023$\\
$\Omega_{\rm c}h^2$&&$0.1171\pm0.0027$&$0.1170^{+0.0027}_{-0.0028}$&$0.1169^{+0.0028}_{-0.0029}$\\
$100\theta_{\rm MC}$&&$1.04107\pm0.00042$&$1.04107^{+0.00040}_{-0.00045}$&$1.04107^{+0.00043}_{-0.00042}$\\
$\tau$&&$0.063\pm0.012$&$0.060\pm0.014$&$0.059^{+0.015}_{-0.017}$\\
$w$&&$-$&$-1.014\pm0.042$&$-0.99^{+0.09}_{-0.09}$\\
$w_a$&&$-$&$-$&$-0.09^{+0.41}_{-0.32}$\\
$N_{\rm eff}$&&$2.99\pm0.16$&$2.97^{+0.18}_{-0.19}$&$2.95\pm0.19$\\
$n_s$&&$0.9662^{+0.0067}_{-0.0069}$&$0.9644^{+0.0085}_{-0.0087}$&$0.9637\pm0.0087$\\
${\rm{ln}}(10^{10}A_s)$&&$3.052\pm0.023$&$3.046\pm0.029$&$3.044^{+0.031}_{-0.035}$\\
$\Omega_{\rm m}$&&$0.3055^{+0.0067}_{-0.0068}$&$0.3041^{+0.0086}_{-0.0087}$&$0.3041^{+0.0088}_{-0.0095}$\\
$H_0$&&$67.7\pm1.1$&$67.9\pm1.2$&$67.8\pm1.2$\\
$\sigma_8$&&$0.807\pm0.011$&$0.808\pm0.012$&$0.809\pm0.012$\\

\hline
$\chi^{2}_{\rm min}$ &&13 903.20&13 902.14 & 13 901.12\\
\hline
\end{tabular}
\label{tab:overallneff}
\end{table*}

\section*{Acknowledgements}
This work was supported by the National Natural Science Foundation of China (grants no.~11522540 and no.~11690021), the Top-Notch Young Talents Program of China and the Provincial Department of Education of Liaoning (grant no.~L2012087).

\bibliographystyle{mn2e}
\def\aj{AJ}
\def\araa{ARA\&A}
\def\apj{ApJ}
\def\apjl{ApJ}
\def\apjs{ApJS}
\def\ao{Appl.Optics}
\def\apss{Ap\&SS}
\def\aap{A\&A}
\def\aapr{A\&A~Rev.}
\def\aaps{A\&AS}
\def\arnpc{Ann. Rev. Nucl. Part. Sci}
\def\azh{AZh}
\def\baas{BAAS}
\def\epjc{Eur. Phys. J. C}
\def\jcap{J. Cosmol. Astropart. Phys.}
\def\jrasc{JRASC}
\def\memras{MNRAS}
\def\na{New Astronomy}
\def\nat{Nature}
\def\npb{Nucl. Phys. B}
\def\npps{Nucl. Phys. Proc. Suppl.}
\def\mnras{MNRAS}
\def\ppnp{Prog. Part. Nucl. Phys.}
\def\pla{Phys. Lett. A}
\def\plb{Phys. Lett. B}
\def\pra{Phys. Rev. A}
\def\prb{Phys. Rev. B}
\def\prc{Phys. Rev. C}
\def\prd{Phys. Rev. D}
\def\prl{Phys. Rev. Lett.}
\def\pasp{PASP}
\def\pasj{PASJ}
\def\physrep{Phys. Rep.}
\def\qjras{QJRAS}
\def\rp{Rev. Phys.}
\def\skytel{S\&T}
\def\solphys{Solar~Phys.}
\def\sovast{Soviet~Ast.}
\def\ssr{Space~Sci. Rev.}
\def\zap{ZAp}
\let\astap=\aap
\let\apjlett=\apjl
\let\apjsupp=\apjs

\bibliography{mybib}

\end{document}